# Estimating Item Difficulty Using Large Language Models and Tree-Based Machine Learning Algorithms

Pooya Razavi*, Sonya Powers

Edmentum (Bloomington, MN, USA)

* The corresponding author, Pooya Razavi, can be reached at prazavi@umass.edu. (ORCiD: 0000-0001-9878-7873)

## Abstract

Estimating item difficulty through field-testing is often resource-intensive and time-consuming. As such, there is strong motivation to develop methods that can predict item difficulty at scale using only the item content. Large Language Models (LLMs) represent a new frontier for this goal. The present research examines the feasibility of using an LLM to predict item difficulty for K-5 mathematics and reading assessment items ($N = 5170$). Two estimation approaches were implemented: (a) a direct estimation method that prompted the LLM to assign a single difficulty rating to each item, and (b) a feature-based strategy where the LLM extracted multiple cognitive and linguistic features, which were then used in ensemble tree-based models (random forests and gradient boosting) to predict difficulty. Overall, direct LLM estimates showed moderate to strong correlations with true item difficulties. However, their accuracy varied by grade level, often performing worse for early grades. In contrast, the feature-based method yielded stronger predictive accuracy, with correlations as high as $r = .87$ and lower error estimates compared to both direct LLM predictions and baseline regressors. These findings highlight the promise of LLMs in streamlining item development and reducing reliance on extensive field testing and underscore the importance of structured feature extraction. We provide a seven-step workflow for testing professionals who would want to implement a similar item difficulty estimation approach with their item pool.

## Estimating Item Difficulty Using Large Language Models and Tree-Based Machine Learning Algorithms

Estimating item difficulty through field-testing is often resource-intensive and time-consuming (AlKhuzaey et al., 2024; Veldkamp & Matteucci, 2013). Large-scale assessments usually devote considerable effort to pretesting items to obtain difficulty estimates, which can introduce significant delays between item development and use (Benedetto et al., 2023) and raise concerns about item exposure and test security in high stakes testing (Way, 1998). Manual difficulty ratings by subject matter experts (SME) are sometimes used as an alternative; but such evaluations are time-consuming, and might have insufficient or inconsistent accuracy (Bejar, 1981; Sayın & Bulut, 2024; Štěpánek et al., 2023; Sydorenko, 2011). Given the importance of accurately calibrated item difficulties, for example, to deliver targeted questions in adaptive testing, there is strong motivation to develop automatic methods that can predict item difficulty using only the item content.

During the past decade, researchers have turned to natural language processing (NLP) techniques to estimate item difficulty directly from the text or content of questions (Benedetto et al., 2023). The rationale is that the linguistic and cognitive features of an item (e.g. vocabulary, syntax, conceptual complexity) could be analyzed to predict how difficult the item would be for students, without needing extensive pilot testing. Early approaches in this area used a combination of surface text features and predictive modeling. For example, researchers extracted surface features such as sentence length, word rarity, and grammatical structure from items, then used regression models to predict empirical difficulty indices (Benedetto et al., 2020). Some studies achieved moderate success, but also revealed the limits of simple features. Readability formulas (e.g. Flesch-Kincaid grade level), while intuitive, often failed to distinguish easy vs.

hard questions. Ha et al. (2019), for instance, reported that standard readability metrics were weak predictors of actual item difficulty in a medical licensing exam. These findings suggest that predicting item difficulty may depend on deeper semantic and cognitive factors.

To incorporate richer linguistic information, some researchers have examined using distributed representations and more sophisticated models. Approaches based on word embeddings and predictive modeling showed improved performance in difficulty prediction (Hsu et al., 2018). Ha et al. (2019) built a random forest model using a wide range of linguistic features and embedding types and found that using a combination of these features yielded significantly higher predictive power ($r$ = .32) compared to using single surface level features (e.g., word count [$r$ = .05] or Flesch Reading ease index [$r$ = -.01]). Still, the overall prediction accuracy in early studies was modest, underscoring the challenge of the task. Furthermore, while many studies have demonstrated the capabilities of predicting item difficulty using item features and embeddings for contexts such as language learning, results have been mixed for efforts to generalize these approaches beyond these subject areas (AlKhuzaey et al., 2024; El Masri et al., 2017; Yaneva et al., 2024).

Large Language Models (LLMs) such as the GPT series represent a new frontier for addressing this problem. LLMs are transformer-based models trained on massive text corpora, enabling them to capture nuanced linguistic patterns, world knowledge, and even some reasoning abilities. These models have achieved remarkable success on various language tasks, including question answering and educational benchmarks. In the context of item difficulty, LLMs offer several potential advantages. First, they can encode linguistic features beyond what previous automated text analysis methods cover (e.g., Rathje et al., 2024). Second, LLMs possess substantial world knowledge and conceptual understanding that could enhance their ability to evaluate item difficulty. For example, a GPT model might recognize that a math problem involving multiple steps or a less familiar concept is likely harder for elementary students than a single-step addition problem. Third, LLMs can arguably perform reasoning and simulation (Havrilla et al., 2024): by attempting to solve or analyze an item in context, an LLM might gauge the cognitive processes required and thereby infer difficulty.

**Related Work**

Recent research has explored two primary paradigms for leveraging LLMs in item difficulty estimation[1]: treating LLMs as proxies for human examinees and using LLMs as direct feature extractors or parameter estimators (see Table 1 for an overview).

The first paradigm, "AI as examinee," involves using language models to simulate student responses (Benedetto et al, 2024; Liu et al., 2025; Maeda, 2025; Martynova et al., 2025; Park et al., 2024; Uto et al, 2024). An example of this concept comes from Uto et al. (2024), who used multiple question-answering (QA) systems as "artificial students." By applying Item Response Theory (IRT) and test-linking procedures, they mapped the QA systems' performance onto a human-calibrated scale, achieving moderate correlations between predicted and ground-truth item difficulty ($r \approx .33$–$.59$) for English vocabulary and reading comprehension tests. More recently, researchers have applied this idea to modern LLMs. Benedetto et al. (2024) used prompt engineering with models like GPT-3.5 and GPT-4 to simulate students of varying proficiency in Science and English reading comprehension tests. The overall patterns in the simulated performances indicated that LLMs have the potential to be used as QA systems; however, the authors cautioned that the difficulty estimates from the models were not reliable and “cannot be directly used as an indication of question difficulty” (p. 11356). Park et al. (2024) proposed the *LLMs are Students at Various Levels (*LLaSA) framework, which models clusters of diverse LLMs as students at different proficiency levels. Their zero-shot LLaSA variant achieved a correlation of $r$ = .35 with IRT-calibrated difficulties. Furthering this line of inquiry, Maeda (2025) fine-tuned multiple transformer models to act as "AI examinees" for English grammar items. While this approach was successful at reproducing the probability of correct response curves ($r$ ≈ .74–.82), it yielded weak-to-modest correlations with calibrated difficulty parameters generated from human examinee responses ($r$ ≈ .08–.22 for full data). Collectively, these studies show that while simulating test-takers is a promising avenue, accurately recovering latent difficulty parameters remains a challenge.

[1] While a comprehensive survey of all recent research on text-based approaches to item difficulty estimation is beyond the scope of the present manuscript, readers interested in a broader synthesis of simulation-based and modeling-based paradigms in this rapidly evolving area are referred to Peters et al. (2025), which provides an extensive overview of the literature.

The second paradigm treats LLMs as direct estimators, using their capabilities to analyze item content and predict difficulty. Early work in this area demonstrated the potential of transformer models; for instance, Zhou and Tao (2020) trained a BERT-based model to classify the difficulty of programming questions, achieving approximately 67% accuracy. A more integrated psychometric approach was taken by McCarthy et al. (2021), who used pretrained BERT embeddings as passage-level features within a linear logistic test model (LLTM) to predict item parameters for an English language proficiency test. Their model produced moderate-to-strong correlations between predicted and observed item mean scores ($r$ ≈ .52–.59) for new items, demonstrating that transformer-based features can effectively "jump-start" the calibration process. These studies are part of a broader trend where transformer models, when appropriately adapted, typically outperform models that rely solely on handcrafted linguistic features (AlKhuzaey et al., 2024).

**Table 1.**

*Summary of Prior AIML Approaches to Item Difficulty Estimation (As Discussed in the Introduction)*

| Study | Domain | Method | Reported Performance | Comments |
|---|---|---|---|---|
| Ha et al. (2019) | Medical licensing exam | Surface + linguistic features + Random Forest | $r = .32$ | Traditional NLP feature engineering; modest predictive performance |
| McCarthy et al. (2021) | English language proficiency | BERT embeddings + LLTM | $r \approx .52$–.59 | Transformer embeddings used within a psychometric modeling framework |
| Zhou & Tao (2020) | Programming questions | BERT-based classification | ~67% accuracy (classification) | Focused on difficulty classification rather than continuous IRT parameter recovery |
| Uto et al. (2024) | Vocabulary & Reading | Multiple QA systems as "artificial students" + IRT linking | $r \approx .33$–.59 | "AI as examinee" paradigm |
| Benedetto et al. (2024) | Science & Reading comprehension | GPT-3.5 role-playing students at 5 skill levels (zero-shot) | $r = 0.13$ (virtual vs. human difficulty) | Successful skill-level simulation; weak difficulty recovery; prompts model-specific |
| Park et al. (2024) (LLaSA) | Mixed academic domains | Clustered LLM simulation (zero-shot) | $r = .35$ | Models LLM clusters as proficiency levels |
| Maeda (2025) | English grammar | Fine-tuned transformer "AI examinees" | $r \approx .08$–.22 (difficulty); $r \approx .74$–.82 (response curve) | Strong fit to response curves; modest recovery of calibrated difficulty |
| Yaneva et al. (2024) | Clinical MCQs | Multi-team modeling comparison | RMSE gain ≤ 0.012 relative to baseline | Demonstrates difficulty of surpassing baseline models |

*Note.* The goal of the table is to offer readers a concise comparison of methodological paradigms covered in the Introduction. For a comprehensive survey of text-based difficulty estimation methods, refer to Peters et al. (2025).

Considering the vast world knowledge of LLMs and their ability to interpret content in intricate ways, a key question is the extent to which a zero-shot approach—harnessing the model's own internal knowledge without task-specific training—could yield reasonable difficulty estimates. Could we instruct an LLM, via well-informed natural language prompting, to judge how difficult a test item would be for students? Furthermore, could we leverage the higher-level capabilities of LLMs as expert judges to rate different features of items based on their content, which we can then use for predicting item difficulty? If feasible, these approaches might provide a fast and cost-effective way to get difficulty estimates, complementing or even expediting the traditional calibration process. Moreover, by focusing on deeper conceptual features rather than superficial text characteristics, these methods may offer greater generalizability to subjects such as mathematics, where traditional NLP features have shown limited success.

**Present Research**

In this study, we investigate the use of GPT-based large language models to estimate the difficulty of K-5 mathematics and reading assessment items. The 'true' difficulty values we aim to predict were established using Item Response Theory (IRT), a widely used psychometric framework for modeling the relationship between a person's latent ability and their responses to assessment items. Specifically, we used item parameters estimated under the Rasch model (Rasch, 1960), a one-parameter IRT model. In this framework, an item's difficulty parameter, denoted as '$b$', represents the ability level ('$\theta$') at which a student has a 50% probability of answering the item correctly; higher '$b$'-values signify more difficult items. For the purposes of this study, these empirically derived Rasch difficulty estimates serve as the ground truth against which we evaluate our model predictions. The items were sourced from Edmentum's Exact Path Diagnostic, a commercial K–12 adaptive assessment, and were previously calibrated by psychometricians using data from large and diverse student populations (further details are provided in the Method section).

Item difficulty prediction in grades K-5 poses unique challenges (young learners' skills are rapidly developing, and small differences in phrasing or context can greatly affect difficulty), making this grade range an ideal target for evaluating LLM's capabilities and limitations. Building on prior research, we employ two approaches to leverage GPT for item difficulty estimation.

First, a direct LLM estimation approach is used: we craft a detailed prompt that describes the task and asks the model (GPT-4o) to analyze a given item's content and predict its difficulty on a numerical scale. We then apply a simple linear transformation trained on a subset of the data to align the LLM's scale with the IRT scale. This approach treats the LLM as an expert evaluator that generates a difficulty rating based on its internal knowledge of language and common curricular expectations. Notably, this is done in a zero-shot manner—the model is not fine-tuned on any item data, but rather instructed to simulate the role of an educational expert. We examine the estimation error of GPT's predicted difficulty scores using root mean square error (RMSE) and mean absolute error (MAE), and the correlations between predicted and "true" difficulties derived from IRT calibration using data collected from actual examinees.

Second, a feature-based modeling approach is implemented, in which we instruct GPT to extract interpretable features from each item. The features are determined based on extensive discussions with math and reading subject matter experts (SMEs) as well as the prior literature on the item characteristics that correlate with item difficulty. We then instruct the LLM to rate each item based on these features. The LLM-generated feature ratings together with item meta-data sourced from the Diagnostic's item pool (e.g., subject domain, item word count) are entered as predictors in ensemble tree-based machine learning models (specifically, random forests and gradient boosting machines), which are trained to predict item difficulty. This feature-based method is structurally in line with previous text-based difficulty modeling approaches discussed earlier; however, the use of a LLM allowed us to extract abstract features in ways that might not be feasible for simpler text-based methods (e.g., dictionary-based approaches). The same set of items with known difficulty values were used for both methods.

These approaches, if successful, offer scalable, cost-effective alternatives to traditional methods that often rely on extensive pilot testing and psychometric analysis. By leveraging rich linguistic features and patterns in item content, such models can support more efficient item development cycles, facilitate pre-screening of items for potential difficulty levels, and ultimately contribute to the creation of more balanced and equitable assessments. As the field continues to explore the integration of artificial intelligence into educational research and practice, understanding the potential and limitations of these methods is both timely and necessary.

## Method

### Items

The items for this study were sourced from Edmentum's Exact Path Diagnostic (henceforth, the Diagnostic), a widely used K-12 computer-adaptive assessment designed to identify student learning gaps and generate personalized learning paths. All Diagnostic items are calibrated within the Rasch IRT framework by psychometricians prior to becoming operational. The calibration for newly developed items is conducted regularly using a sample of 750 students for each item, comprising 250 students from the target grade, 250 from one grade below, and 250 from one grade above. This calibration data is drawn from the large and diverse pool of Diagnostic users (He et al., 2024; Razavi et al., 2024). The Diagnostic's external validity has been demonstrated through linking and correlation studies indicating a strong association between its results and various state (e.g., He et al., 2024; Razavi et al., 2024) and national assessments (e.g., Eckerly et al., 2025). Furthermore, extensive Differential Item Functioning (DIF) research has shown the robustness and fairness of the Diagnostic's items for students from different demographic backgrounds (EdMetric, 2022; eMetric, 2025).

Our analyses were conducted on a total of 5,170 items ($n_{\text{math}}$ = 2564; $n_{\text{reading}}$ = 2606) covering grades K through 5. The selection of these specific items from the larger Exact Path item bank was guided by several key considerations. First, to ensure sufficient data for model training and validation, we targeted a sample size of 300 to 400 items per grade. This target was achieved for all grades except Kindergarten (where item pools are typically smaller), with final sample sizes ranging from 293 to 625 items per grade (see Table 2). Second, to ensure the findings are relevant for ongoing item development, we prioritized items developed and calibrated within the two years preceding this study. Finally, the selection was curated to ensure a reasonable representation of different content subdomains. The item difficulty parameters for this set, previously calibrated within the Rasch IRT framework, ranged from -5.52 to 3.68 for mathematics and -6.33 to 4.03 for reading. Note that the Diagnostic item pool is vertically scaled, meaning the IRT difficulty parameters for items across all grade levels are placed onto a single, continuous scale. This framework allows for the direct comparison of item difficulties and enables the measurement of student proficiency growth across grades.

In addition to the item content and IRT difficulties, the Exact Path Diagnostic pool also contains meta-data for each item, including word count (i.e., total number of words in the item, including the response options), item type (or format), and subject domains. The selected item set was composed predominantly of multiple-choice questions (85.4%), with the remainder comprising various technology-enhanced formats, including fill-in-the-blank (5.2%), graphic gap match (3.4%), hot spot (2.0%), and other types (4.0%). For each subject, the items covered a diverse range of domains; in mathematics, these included Numbers & Operations (26.4%), Measurement, Data, & Statistics (21.3%), Algebra & Expressions (20.7%), Fractions & Ratios (14.9%), Geometry (12.8%), and Counting & Cardinality (3.9%), while in reading, the domains were Reading Literature (34.1%), Reading Informational Text (27.5%), Language and Vocabulary (21.5%), and Reading Foundations (16.9%).

A robust model evaluation requires that the holdout sample accurately reflects the characteristics of the training data, particularly concerning the outcome variable—in this case, item difficulty. A simple random split could result in dissimilar difficulty distributions between the training and testing sets (e.g., a training set with predominantly easy items), which would risk overfitting and limit the model's generalizability across the full difficulty spectrum. To mitigate this, we created the holdout sample using a stratified sampling approach based on the Rasch difficulty estimates. For each subject, we designated 600 items (approximately 23% of each subject's items) for the holdout set. This was implemented using the *caret* R-package (Kuhn, 2020). This procedure ensured that the distribution of item difficulty in the training set was highly similar to that of the holdout set, providing a reliable basis for model validation. In total, there are 3970 training and 1200 holdout items. The final counts and average difficulty estimates for each subject and grade combination are presented in Table 2.

**Table 2.**
*Summary of Math and Reading Item Counts and Parameters*

| | | Math | | | | Reading | | | |
|---|---|---|---|---|---|---|---|---|---|
| | | | Difficulty Estimate | | | | Difficulty Estimate | | |
| Grade | Subset | Count | Min. | Mean | Max. | Count | Min. | Mean | Max. |
| K | train | 220 | -5.32 | -3.49 | -1.45 | 324 | -6.33 | -2.35 | -0.55 |
| | test | 73 | -5.16 | -3.53 | -2.30 | 100 | -5.67 | -2.35 | -0.71 |
| 1 | train | 263 | -5.14 | -3.02 | -0.14 | 306 | -4.31 | -1.74 | 0.68 |
| | test | 100 | -5.52 | -3.11 | -0.69 | 100 | -4.33 | -1.73 | 0.08 |
| 2 | train | 300 | -4.96 | -2.46 | -0.31 | 300 | -4.37 | -1.28 | 1.24 |
| | test | 100 | -4.73 | -2.27 | -0.60 | 100 | -3.84 | -1.51 | 1.17 |
| 3 | train | 382 | -3.54 | -1.16 | 3.68 | 395 | -3.65 | -0.16 | 3.02 |
| | test | 100 | -3.75 | -1.21 | 1.37 | 100 | -3.40 | -0.21 | 2.28 |
| 4 | train | 498 | -5.06 | -0.62 | 2.23 | 381 | -3.32 | 0.24 | 3.53 |
| | test | 127 | -4.49 | -0.54 | 2.36 | 100 | -2.04 | 0.24 | 3.07 |
| 5 | train | 301 | -3.18 | 0.02 | 3.55 | 300 | -2.17 | 0.39 | 4.03 |
| | test | 100 | -2.50 | 0.15 | 3.10 | 100 | -2.02 | 0.30 | 2.56 |

*Note*. Difficulty estimates are in Rasch logit units.

**LLM**

We used OpenAI's GPT-4o model (2024-11-20 version)[2], which is an optimized version of the GPT-4 architecture. GPT-4o was trained by OpenAI on a mixture of publicly available and licensed data, including a wide range of internet text sources. We accessed GPT-4o via OpenAI's API using the *openai* library in Python. To increase consistency and replicability, we controlled for output variability by setting the temperature parameter to zero for all API calls.

[2] Prior to finalizing our choice of GPT-4o, we tested four models (GPT-4o-mini, GPT-4o, Llama 3.2, and Claude-3-haiku) on a subset of items ($n$ = 250). We tested the same zero-shot prompt with all models instructing them to evaluate item difficulties based on item content. We chose GPT-4o due to its better performance based on a comparison of RMSEs and the correlations between predicted and true difficulty estimates.

**Feature Elicitation and Prompt Development**

To ensure our prompts, and the features included in them, were both theoretically grounded and practically relevant to the specific item pool under study, we implemented a structured process involving Subject Matter Experts (SMEs). The SMEs who participated in this study were four members of the in-house assessment design team responsible for the Exact Path Diagnostic. All SMEs had a minimum of five years of experience in K-12 assessment design in their respective subjects (one expert covered both math and reading, one focused on math, and two focused on reading).

We conducted two one-hour focus group sessions via video conference: a dyad for mathematics and a triad for reading. To structure the discussion, SMEs were provided with a set of guiding questions two days prior to the session. The questions were designed to elicit detailed information about the factors influencing item difficulty:

(a) What is the item development process from initial concept to final calibration?

(b) How do you intentionally design item characteristics to target different difficulty levels?

(c) When presented with an item, what characteristics would you examine to estimate its difficulty?

(d) Based on your experience with this specific item pool, what features consistently make items easier or more difficult for students?

The focus group sessions were transcribed. We then analyzed the transcripts to identify and catalog all item characteristics mentioned as potential drivers of difficulty. This SME-derived list was then cross-referenced with features identified in the literature cited in this paper's Introduction to ensure a comprehensive final set. It is important to clarify that our intent was not to replicate the exact quantitative metrics found in prior literature, such as traditional readability scores. Rather, our cross-referencing with the literature was designed to ensure conceptual alignment (e.g., covering constructs like linguistic complexity or cognitive load) instead of replicating specific, often rule-based, metrics from earlier studies (e.g., traditional readability formulas). Our approach relies on the LLM's advanced capacity to perform a more holistic, human-like evaluation of these constructs, moving beyond the limitations of older quantitative proxies. We posit that state-of-the-art LLMs can evaluate abstract qualities like text complexity or cognitive demand in a way that more closely mirrors human expert judgment, thus obviating the need for more rigid, formulaic proxies used in past research.

The complete list of features for both subjects is provided in Table 3. A key insight from this process was the identification of both domain-general features, which SMEs considered relevant to both subjects (e.g., Cognitive Load, Depth of Knowledge), and domain-specific features (e.g., Symbolic Reasoning for math; Figurative Language for reading). This approach of using expert elicitation grounded in a specific item pool generated a rich set of features with a modest time investment from the SMEs (approximately two hours each) and increased our confidence that the selected features were directly relevant to the predictive task.

**Table 3.**

*Item Features for LLM Evaluation, Derived from SME Focus Groups*

| Subject | Feature | Description | Scale |
|---|---|---|---|
| Math | Skill Challenge | The difficulty or advanced nature of the mathematical skill being assessed. | 1 to 10 |
| | Text Translation | The extent to which a student must convert written text into a mathematical operation. | 1 to 10 |
| | Distractor Trickiness | The difficulty and plausibility of the incorrect answer choices. | 1 to 10 |
| | Cognitive Load | The mental effort required, considering working memory and executive function. | 1 to 10 |
| | Depth of Knowledge (DOK) | The complexity of thinking required, based on Webb's DOK framework. | 1 to 4 |
| | Linguistic Cues | The likelihood that students might misinterpret the language or phrasing in the item. | 1 to 10 |
| | Real-World Application | Whether the item requires applying math concepts to a realistic scenario. | Y / N |
| | Improper/Mixed Fractions | Whether the item involves improper fractions or mixed numbers. | Y / N |
| | Visual Information | Whether the item requires interpreting information from visuals like graphs or tables. | Y / N |
| | Multi-Step Reasoning | Whether solving the item requires a logical sequence of more than one step. | Y / N |
| | Concept Integration | Whether the item involves combining multiple mathematical concepts. | Y / N |
| | Evaluate Reasoning | Whether the item asks the student to assess another's mathematical work or logic. | Y / N |
| | Multiple Calculations | Whether solving requires several mathematically distinct calculations or intermediate steps to reach the final answer. | Y / N |
| | Student Engagement | How interesting or engaging the item is for a typical student at that grade level. | 1 to 10 |
| | Symbolic Reasoning | Whether the item requires using algebraic expressions or variables. | Y / N |
| | Math-Specific Terms | Whether the item uses formal mathematical vocabulary. | Y / N |
| | Multi-Part Instructions | Whether the item's instructions consist of multiple distinct commands. | Y / N |
| | Repeated Calculation-Subconclusion | Whether solving requires applying the same calculation and making sub-conclusions multiple times. | Y / N |
| | Conceptual vs. Arithmetic | The primary demand of the item: conceptual understanding, arithmetic, or both. | 1 to 4 |
| | Overall Difficulty | A holistic, final estimate of the item's difficulty on a fine-grained scale. | 1 to 100 |
| Reading | Skill Challenge | The difficulty or advanced nature of the reading skill being assessed. | 1 to 10 |

| | | |
|---|---|---|
| Vocabulary Complexity | The complexity or advancement of the vocabulary used in the item and passage. | 1 to 10 |
| Distractor Trickiness | The difficulty and plausibility of the incorrect answer choices. | 1 to 10 |
| Cognitive Load | The mental effort required, considering memory, executive function, and metacognition. | 1 to 10 |
| Depth of Knowledge (DOK) | The complexity of thinking required, based on Webb's DOK framework. | 1 to 4 |
| Inference Required | Whether the item requires students to infer information not explicitly stated. | Y / N |
| Multiple Constructs | Whether the item assesses more than one distinct reading skill or concept. | Y / N |
| Syntax Complexity | The grammatical complexity of the sentences in the item and passage. | 1 to 10 |
| Figurative Language | The extent to which abstract or figurative language is used. | 1 to 10 |
| Student Engagement | How interesting or engaging the item is for a typical student at that grade level. | 1 to 10 |
| Passage Difficulty | A rating of the reading difficulty of the associated passage, if applicable. | 1 to 10 or N/A |
| Multi-Step Reasoning | Whether solving requires a sequence of multiple steps or distinct lines of reasoning. | Y / N |
| Overall Difficulty | A holistic, final estimate of the item's difficulty on a fine-grained scale. | 1 to 100 |

*Note.* This table presents the features that were developed for evaluation and extraction by the LLM. While SMEs identified other important factors (e.g., grade level, domain, word count, item type), these were already available as item metadata in the test bank and were therefore excluded from the LLM rating task.

## Difficulty Estimation Approaches

Grounded in the features identified in the preceding section, we developed two distinct approaches to leverage the LLM for item difficulty estimation: (1) a direct estimation approach and (2) a feature-based modeling approach. The prompts and procedures for each are described below.

### *Direct LLM Estimation*

Using a zero-shot learning approach (Schulhoff et al., 2024), we created a detailed prompt that asks the model (GPT-4o) to analyze a given item's content and predict its difficulty on a scale from 1 to 100[3]. The final prompt (presented in Appendix A) provides the model with item content (e.g., item prompt, stem, and response options) as well as meta data (e.g., grade level and item type [e.g., multiple-choice, fill-in-the-blank]). It instructs the language model to act as an expert in K–12 math or reading assessment and evaluate the difficulty of a given item based on its content and metadata. It guides the model to reason step-by-step using factors such as the grade level, item type, skill complexity, distractor quality, cognitive load, and Depth of Knowledge (DOK; Webb, 1997) level. The prompt also includes specific features (identified using the insights from the discussions with SMEs) that may influence item difficulty, such as multi-step reasoning, use of visuals, or linguistic complexity. The model is then asked to synthesize this information and provide a numerical difficulty rating on a scale from 1 (very easy) to 100 (very challenging), using defined difficulty bands.

[3] Before finalizing the prompt, we conducted an iterative refinement process using a subset of items ($n$ = 250). This process involved modifying the prompt to test different difficulty estimation ranges (e.g., -3 to +3, -5 to +5, and 1 to 100), as well as revising the evaluation criteria included in the instructions. Iterations continued until improvements in model performance (measured by RMSE and the correlation between predicted and true difficulty estimates) began to plateau. We also examined few-shot learning prompts, where the model was provided with examples of items with relatively high and low difficulty estimates. The few-shot learning approach did not produce improved results.

After the LLM-based estimates are generated, to put them on the same scale as the Rasch logit difficulty estimates, we first performed a standard score transformation (z-score), then rescaled the standardized estimates to match the mean and standard deviation of the true Rasch logit estimates. The equation below describes the rescaling process:

$$GPT_{estimate\ rescaled_i} = \left(\frac{\left(GPT_{raw\ estimate_i} - \mu_{GPT}\right)}{\sigma_{GPT}}\right) \times \sigma_{Rasch} + \mu_{Rasch}$$

Where:

$GPT_{raw\ estimate}$ represents the original LLM estimate for item *i* (on the 1-100 scale).

$\mu_{GPT}$ and $\sigma_{GPT}$ represent the mean and standard deviation of the LLM raw estimates.

$\mu_{Rasch}$ and $\sigma_{Rasch}$ represent the mean and standard deviation of the Rasch logit estimates.

$GPT_{estimate\ rescaled\,i}$ represents the final rescaled estimate for item *i*.

We then fit a regression model on the training dataset for each subject and grade where the true item difficulty estimates are the outcome variable and the rescaled GPT estimates are the predictors:

$$\hat{y}_i = \beta_0 + \beta_1 \cdot GPT_{estimate\ rescaled_i} + \varepsilon_i$$

Since the primary goal of this research is to develop an approach to predict difficulty estimates based on item content alone, we used the parameters from the trained regression models to generate difficulty estimates for the holdout sample. This validation step will give us an understanding of how well this approach would perform on “unseen” data. The model predictions will be examined using evaluation metrics described in a later subsection of the manuscript.

### *Feature-Based Estimation*

In this approach, we used the LLM to extract specific features (Table 3) from each item, as determined based on the focus groups with math and reading SMEs. The prompts (see Appendix B) instruct the language model to act as an expert in K-12 math or reading assessment design, to review the item content and metadata, and to analyze items using a structured approach. For math, the model is asked to answer 20 targeted questions that assess various dimensions of item complexity and cognitive demand. These include skill difficulty, the need for translation from text to math, distractor quality, cognitive load, DOK level, linguistic complexity, use of visuals, multi-step reasoning, the need for integration of concepts, and more. For reading, the model is asked to respond to 13 targeted questions assessing various dimensions of item difficulty and complexity. These include skill challenge, vocabulary and syntax complexity, distractor quality, cognitive load, DOK, use of inference or abstract language, multi-construct assessment, and more. For both subjects, the model is also asked to generate an overall difficulty estimate (on a 1-100 scale) considering all the relevant factors. For each feature, the expected response options or response range are provided to the model. Depending on the feature, the model is asked to provide numeric or binary (Y/N) responses (see Table 3 for feature response scales). Prior to using the LLM responses, we reviewed their variability and excluded one math feature (i.e., "does the item require the respondent to evaluate someone else's calculations?") from further analysis because it had near-zero variance.

We then used the LLM-generated features together with item meta-data (e.g., word count) as predictors in two tree-based machine learning algorithms (i.e., random forests and gradient boosting machines) trained to predict item difficulty estimates.

Random forests and gradient boosting machines are both ensemble learning methods that build multiple decision trees to improve prediction accuracy (James et al., 2021). Random forests operate by constructing a large number of decision trees using bootstrapped samples of the data and averaging their predictions to reduce variance and prevent overfitting. In contrast, gradient boosting builds trees sequentially, with each new tree learning to correct the errors made by the ensemble so far, thereby minimizing bias. While random forests emphasize robustness through randomization and averaging, gradient boosting focuses on model refinement through iterative optimization. Both methods are well-suited to handle nonlinear relationships, interactions among features, and mixed data types, making them strong candidates for modeling item difficulty based on complex, high-dimensional feature sets.

For each subject, we trained a random forest regression model using the *randomForest* package in R (Breiman et al., 2018). The initial model was constructed with 500 trees on the training dataset. To optimize model performance, we implemented 5-fold cross-validation using the *caret* package (Kuhn, 2020). The hyperparameter mtry (number of variables randomly sampled at each split) was systematically tuned by testing values ranging from 2 to the square root of the total number of predictors. A predefined grid search approach was employed for hyperparameter optimization, with RMSE serving as the evaluation metric. The final optimized random forest model retained the 500-tree structure while incorporating the optimal mtry value identified through cross-validation.

Additionally, for each subject we trained a gradient boosting machine (GBM) model using the XGBoost algorithm via the *caret* package in R. The model was systematically optimized through 5-fold cross-validation to ensure robust performance evaluation. A comprehensive hyperparameter grid search was conducted across seven key parameters: number of boosting rounds (100, 200), maximum tree depth (3, 6), learning rate (0.01, 0.1), minimum loss reduction (0, 1), column subsampling ratio (0.8, 1), minimum child weight (1, 5), and instance subsampling ratio (0.8, 1). This resulted in 128 unique model configurations being evaluated. RMSE served as the primary evaluation metric for model selection. Following training, the hyperparameter configuration yielding the lowest RMSE was identified and selected as the optimal model.

**Performance Evaluation Metrics**

To evaluate the performance of both the direct and feature-based approaches, we assess their predictions against two primary criteria: estimation error and correlation. For estimation error, we use Root Mean Square Error (RMSE) and Mean Absolute Error (MAE). In addition to these error-based comparisons, we evaluate the Pearson correlation between the true and predicted difficulty estimates to measure the linear relationship captured by each model. A meaningful model should not only produce lower error than the grade-specific dummy regressor but also show a strong positive correlation with the true difficulty values.

**Performance Benchmarks**

To evaluate the added value of our methods, we used one common benchmark across both estimation approaches and two additional comparison models specific to the feature-based pipeline. For both the direct LLM estimation approach and the feature-based approach, we used a grade-specific dummy regressor as the primary benchmark. This benchmark establishes a fair baseline that any content-based method should exceed, because it uses only grade-level mean difficulty and does not incorporate information from item content.

For the feature-based approach, we additionally included two comparison models: a TF-IDF + random forest model and metadata-only models. We treated these as feature-based comparison models rather than common benchmarks across approaches because they address a different methodological question than the direct LLM approach. Specifically, the direct approach asks whether an LLM can generate a useful holistic difficulty judgment in a zero-shot manner, whereas the feature-based approach asks whether LLM-extracted cognitive and linguistic features improve prediction relative to more conventional structured predictors within a supervised modeling framework. The TF-IDF baseline therefore tests whether LLM-derived features improve upon traditional text-based representations, and the metadata-only models test whether they add predictive value beyond readily available item characteristics such as grade, domain, item type, and word count. Each baseline is described in more detail below.

**Dummy Regressor.** The dummy regressor operates by calculating the mean of the "true" Rasch difficulty values for each grade level separately, using only the training dataset (e.g., see Yaneva et al., 2024). It then uses this corresponding grade-specific mean as its prediction for every item in the test set. Given that our items are vertically scaled, this grade-stratified baseline sets an appropriate reference point. It leverages a key piece of metadata—the grade level—which is inherently correlated with item difficulty. Therefore, for our content-based models to be considered effective, they must demonstrate that they can predict patterns from the item content itself that go beyond the information already provided by an item's grade designation.

**TF-IDF + Random Forest.** This baseline represents traditional natural language processing approaches to item difficulty prediction prior to the widespread adoption of large language models (Ha et al., 2019). We extracted Term Frequency-Inverse Document Frequency (TF-IDF) features from each item's full text content, including the prompt, stem, passages, and response options. Specifically, we generated unigram and bigram features with a maximum vocabulary size of 5,000 terms, excluding terms that appeared in fewer than 2 items or in more than 60% of items to reduce noise and overfitting. We included grade level as an explicit predictor alongside the TF-IDF features to create a fair comparison with our LLM-based approaches, which also have access to this metadata. The random forest model was trained with 800 trees, and hyperparameters (*mtry* and *min.node.size*) were optimized via 5-fold cross-validation on the training set, using RMSE as the selection criterion. The optimal model was then applied to the holdout set for final evaluation. This baseline allows us to assess whether the semantic understanding and reasoning capabilities of LLMs offer improvements over keyword-based text analysis.

**Metadata-Only Models.** To isolate the specific contribution of LLM-extracted features, we fit additional baseline models that used only item metadata as predictors—specifically, grade level, subject domain, item type, and word count—excluding all LLM-rated cognitive and linguistic features. These metadata-only models were implemented separately for mathematics and reading using the same algorithms (random forest and gradient boosting) and identical training procedures (hyperparameter tuning, cross-validation, train/holdout splits) as the full feature-based models. Comparing the metadata-only models with the full models (metadata + LLM-extracted features) allows us to quantify the predictive value added by LLM-rated features beyond what is captured by readily available item characteristics.

## Results

### *Direct LLM Estimation*

In this approach, the LLM was asked to assign a numerical difficulty rating on a 1-100 scale. As described in the Method section, to align these LLM-generated estimates with the Rasch logit scale, we applied a z-score transformation to standardize the estimates, then rescaled them to match the mean and standard deviation of the Rasch logit values. Finally, we fit a regression model (separately for each subject and grade) using the rescaled GPT estimates as predictors of the true Rasch item difficulties.

The regression results (Table 4) indicated that, with the exception of math grade 1, GPT estimates significantly predict the true estimates, with some notable variability in the models' predictive performances. For math, the adjusted $R^2$ values of the significant models ranged from .11 for grade 5 to .23 for grade 4. For reading, the $R^2$ varied between .02 (grade 1) to .45 (grade 3). Regression parameters for each model are presented in Table 4.

**Table 4.**

*Regression Parameters Predicting True Difficulty Estimates from Rescaled LLM Estimates*

| Subject | Grade | $\beta_0$ | $\beta_1$ | Omnibus Test | $R^2$ |
|---|---|---|---|---|---|
| Math | K | -2.61 | 0.24 | $F(1,218) = 33.17, p < .001$ | 0.132 |
| | 1 | -2.73 | 0.09 | $F(1,261) = 0.81, p = 0.369$ | 0.003 |
| | 2 | -1.20 | 0.53 | $F(1,298) = 52.29, p < .001$ | 0.149 |
| | 3 | -0.61 | 0.49 | $F(1,380) = 69.21, p < .001$ | 0.154 |
| | 4 | -0.31 | 0.55 | $F(1,496) = 149.29, p < .001$ | 0.231 |
| | 5 | 0.01 | 0.28 | $F(1,299) = 36.86, p < .001$ | 0.110 |
| Reading | K | -1.45 | 0.38 | $F(1,322) = 29.88, p < .001$ | 0.085 |
| | 1 | -1.05 | 0.32 | $F(1,304) = 6.03, p = 0.015$ | 0.019 |
| | 2 | -0.08 | 0.89 | $F(1,298) = 84.57, p < .001$ | 0.221 |
| | 3 | -0.14 | 0.73 | $F(1,393) = 321.75, p < .001$ | 0.450 |
| | 4 | 0.03 | 0.61 | $F(1,379) = 211.65, p < .001$ | 0.358 |
| | 5 | -0.02 | 0.66 | $F(1,298) = 189.92, p < .001$ | 0.389 |

*Note.* The parameters are based on regression models fit on the training dataset for each subject and grade. In each model, the true item difficulty estimates are the outcome variable and the GPT estimates based on the direct LLM estimation approach are the predictors. The GPT estimates have been rescaled prior to being entered into the model (details provided in the Method section).

We then used the parameters from the trained regression models (Table 4) to generate difficulty estimates for the holdout sample. Below, we discuss the results of the predicted difficulty estimates for each subject on the holdout sample. As a reminder, the performance of the models is compared against a baseline regressor which predicts an item's difficulty using the mean Rasch value for its specific grade level.

**Math.** When comparing the predicted and true difficulty estimates across all grades, both error indices (RMSE = 0.91, MAE = 0.72) were lower than the benchmarks (1.01 and 0.81, respectively), indicating that the GPT estimates perform better than dummy regressor models. We also observed a strong correlation between the true and predicted difficulty estimates, when collapsing all grades together ($r = .83$; Figure 1).

When analyzing the results for each grade individually, we see variability in the prediction accuracy across grades (Table 5). The RMSEs ranged between 0.74 (grade K) and 1.00 (grade 4), and MAEs varied between 0.60 (grade K) and 0.78 (grade 3). Importantly, both error indices for grade K were higher than the benchmarks, showing that even though this grade has the lowest RMSE and MAE, the prediction accuracy is worse than a model predicting item difficulties based on grade averages. The error indices for grade 1 are minimally better than the benchmarks, but the indices for the rest of the grades improved upon the dummy regressor estimates by at least 0.08 points. Likewise, there are moderate correlations between the true and predicted difficulty estimates for grades 3 and above, and low correlations for grades K and 1 (Figure 2).

**Reading.** The overall RMSE (0.86) and MAE (0.69) were considerably lower than the benchmarks (1.04 and 0.84, respectively), indicating that using the difficulty estimates generated by the LLM provides better accuracy compared to estimates generated by a dummy regressor. Further, correlation analyses indicated that the predicted and true estimates are highly correlated ($r = .81$; Figure 1).

The results of the analyses by grade indicate considerable variability in the accuracy of the difficulty estimates across grades, though all error estimates are smaller than the benchmarks. As shown in Table 5, grade 3 has the smallest prediction error (RMSE = 0.78 and MAE = 0.59), grade 2 has the largest (RMSE = 0.96 and MAE = 0.79), and grade 1 has error estimates that improve little upon the benchmarks (RMSE and MAE difference = 0.01). Consistently, the correlation estimates (Figure 2) point to strong associations between true and predicted difficulty estimates for grades 3 through 5, moderate associations for grades K and 2, and low association for grade 1.

**Figure 1.**

*Association between LLM-Estimated and True Difficulty Estimates based on the Direct LLM Estimation Approach*

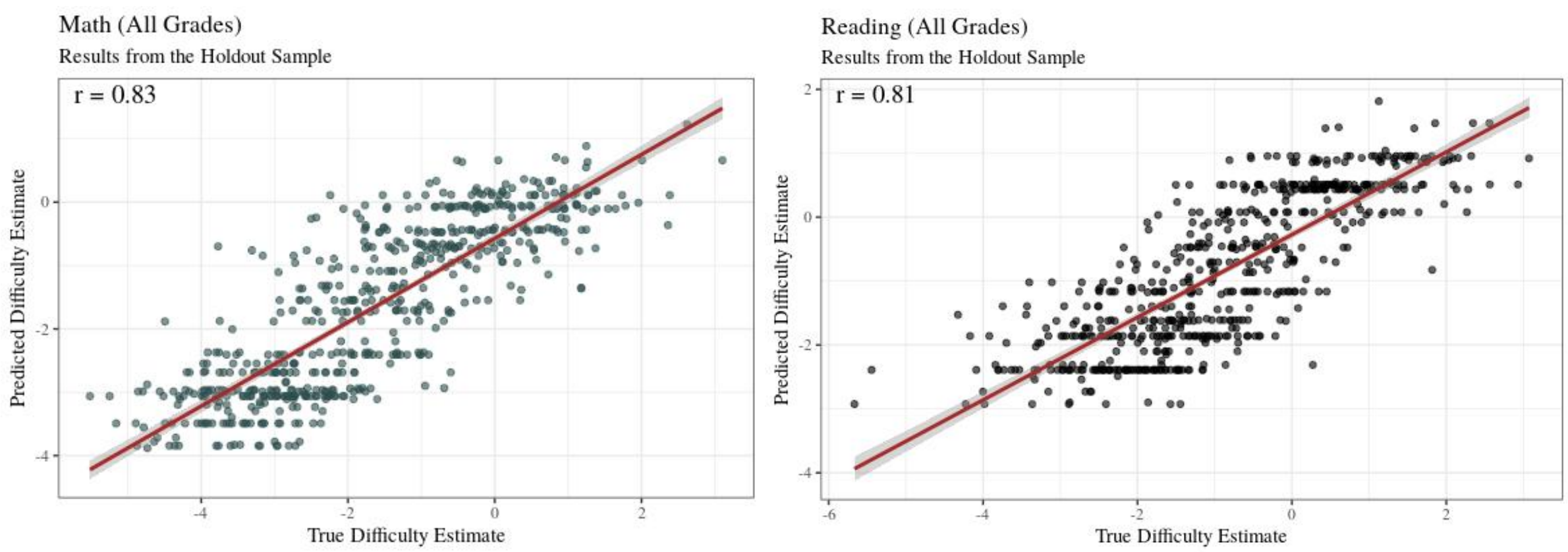

**Table 5.**

*Error Estimates by Subject and Grade based on the Direct LLM Estimation Approach*

| | | RMSE | | | MAE | | |
|---|---|---|---|---|---|---|---|
| Subject | Grade | Dummy | GPT Estimates | Difference | Dummy | GPT Estimates | Difference |
| Math | K | 0.70 | 0.74 | -0.05 | 0.58 | 0.60 | -0.02 |
| | 1 | 0.86 | 0.85 | 0.01 | 0.70 | 0.69 | 0.01 |
| | 2 | 0.96 | 0.84 | 0.12 | 0.82 | 0.71 | 0.11 |
| | 3 | 1.11 | 0.97 | 0.14 | 0.91 | 0.78 | 0.12 |
| | 4 | 1.18 | 1.00 | 0.18 | 0.92 | 0.76 | 0.16 |
| | 5 | 1.05 | 0.95 | 0.10 | 0.83 | 0.75 | 0.08 |
| Reading | K | 0.90 | 0.82 | 0.07 | 0.69 | 0.63 | 0.06 |
| | 1 | 0.90 | 0.89 | 0.01 | 0.74 | 0.73 | 0.01 |
| | 2 | 1.10 | 0.96 | 0.14 | 0.90 | 0.79 | 0.10 |
| | 3 | 1.14 | 0.78 | 0.36 | 0.90 | 0.59 | 0.30 |
| | 4 | 1.06 | 0.82 | 0.24 | 0.84 | 0.64 | 0.19 |
| | 5 | 1.13 | 0.89 | 0.23 | 0.96 | 0.73 | 0.23 |

**Figure 2.**

*Association between LLM-Estimated and True Difficulty Estimates by Grade*

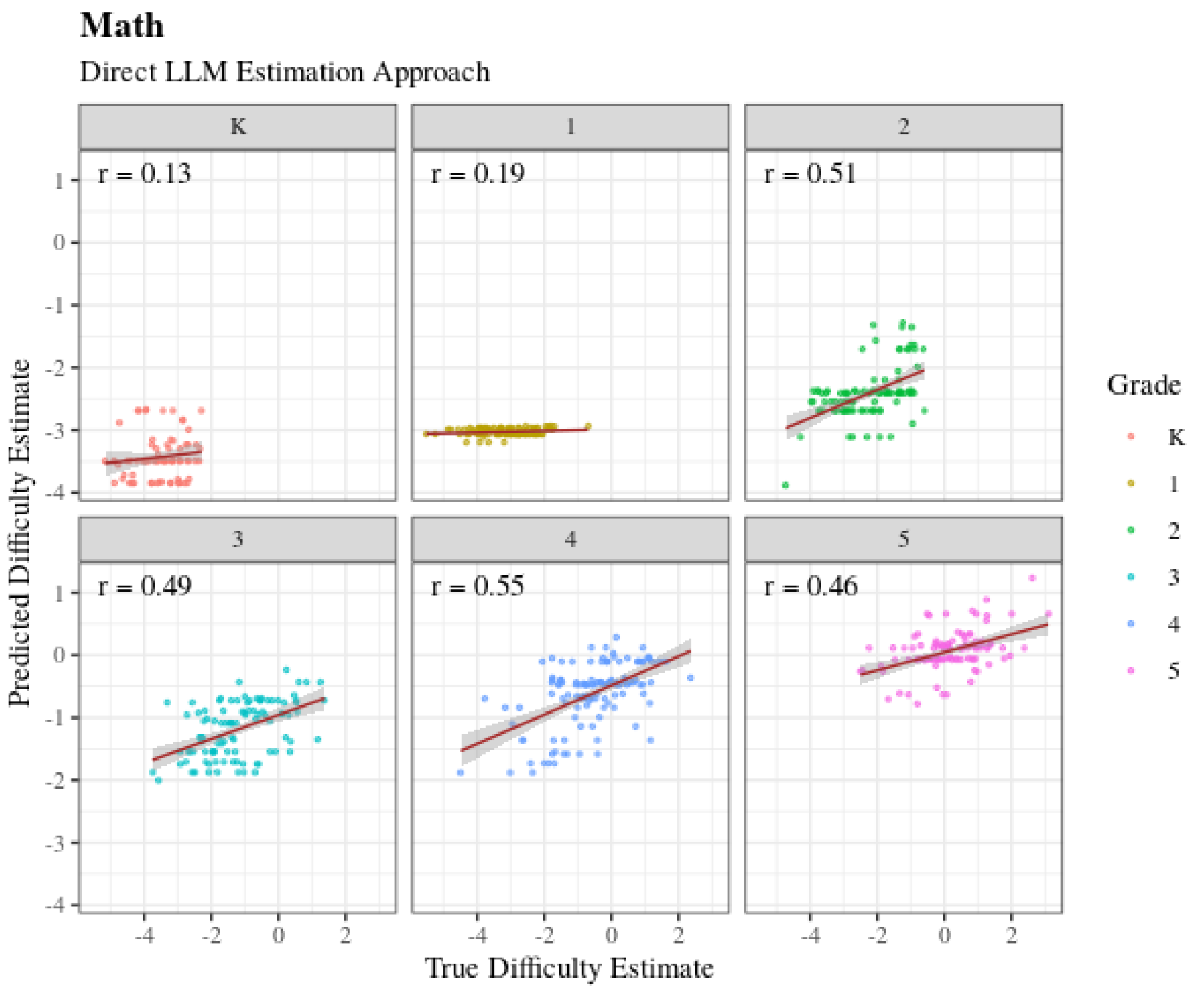

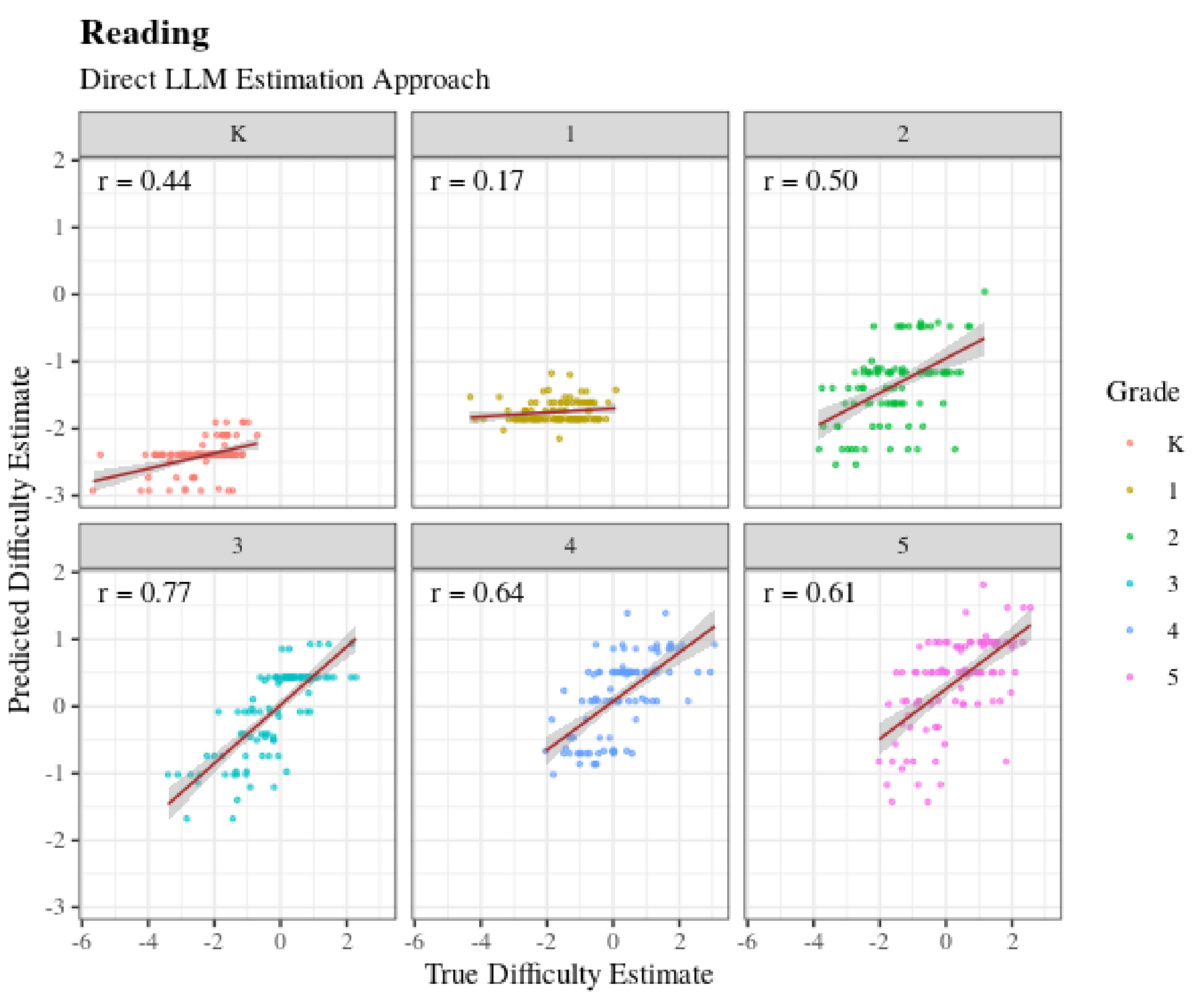

***Feature-Based Estimation***

In this approach, we instructed the LLM to extract a set of features from each item. We then used these features, together with item meta data (i.e., grade, subject, item type, and word count), to train models that predict item difficulty estimates. We used two tree-based machine learning algorithms generally known for their high performance in predictive tasks such as this one: random forests and gradient boosting machines (GBM). After finetuning each model on the training dataset, we used the optimal model to predict item difficulty on the holdout sample. Below we describe the results of the *testing* set (i.e., holdout sample) for each subject based on the two models.

***Math Predictions Using Random Forest.*** The overall comparison of true difficulty estimates and those predicted by the random forest produced an RMSE of 0.83 and MAE of 0.64, showing that the model's predictions are more accurate than the ones from the previous approach (i.e., direct LLM estimation: RMSE = 0.91, MAE = 0.72) and the benchmarks based on the dummy regressor models (RMSE = 1.01, MAE = 0.81), the TF-IDF + Random Forest benchmark (RMSE = 1.01, MAE = 0.80), and the metadata-only random forest model (RMSE = 1.12, MAE = 0.90).The predicted estimates had a strong correlation with the true difficulty estimates ($r$ = .87; Figure 3).

The results for each grade individually point to variability in the prediction accuracy across grades (Table 6). Model performance is noticeably better than the direct LLM estimation approach and the benchmarks for grade 1 through 5. However, similar to the results from the previous approach, the accuracy for grade K is worse than the dummy regressor benchmark. The correlations between true and predicted estimates are, however, consistently higher than those observed in the previous approach ($.37 \leq r \leq .70$; Figure 4). Overall, these results suggest that the combination of LLM feature extraction and predictive modeling using random forest generates more accurate estimates than the direct LLM-estimation approach and the benchmarks, at least for grades 1 through 5.

***Math Predictions Using GBM.*** The overall error estimates based on the predictions generated by GBM were similar to the random forest results (RMSE = 0.81, MAE = 0.63). And the predicted and true difficulty estimates were highly correlated ($r$ = .87; Figure 3).

The by-grade analyses indicated that GBM performs similarly to random forest, with the exception of grade K where the error indices for GBM are better (RMSE = 0.63, MAE = 0.49; Table 6) and the correlations between true and predicted difficulty estimates are higher ($r$ = .53; Figure 4). Overall, across both approaches and all benchmarks, the gradient boosting model emerges as the strongest performer, reducing RMSE compared to the dummy regressor by 18.5% on average across grades, and reducing RMSE relative to the strongest benchmark (i.e., the metadata-only gradient boosting model) by 11.2% on average.

**Figure 3.**

*Association between Feature-Based Estimates and True Difficulty Estimates for Math*

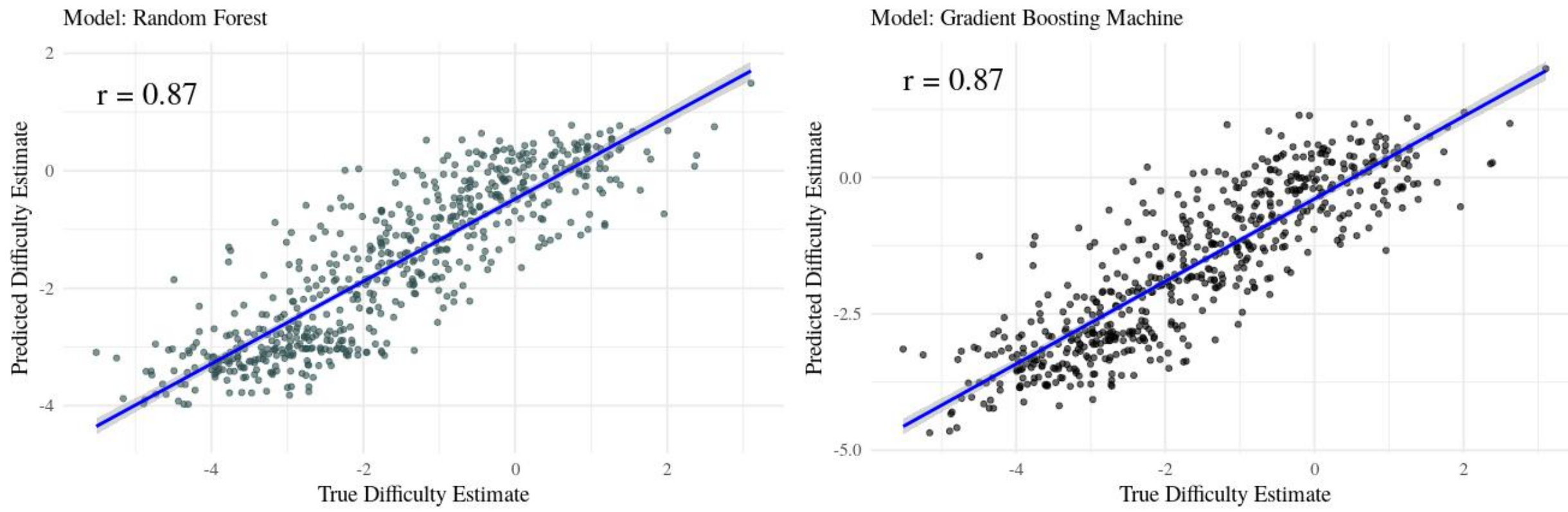

**Table 6.**

*RMSE by Grade Based on the Feature-Based Estimation Approach for Math*

| | Grade | | | | | |
|---|---|---|---|---|---|---|
| Approach | K | 1 | 2 | 3 | 4 | 5 |
| Dummy Regressor | 0.70 | 0.86 | 0.96 | 1.11 | 1.18 | 1.05 |
| Direct LLM Estimation | 0.74<br>(-5.7%) | 0.85<br>(+1.2%) | 0.84<br>(+12.5%) | 0.97<br>(+12.6%) | 1.00<br>(+15.3%) | 0.95<br>(+9.5%) |
| TF-IDF + RF | 0.70<br>(0.0%) | 0.86<br>(-0.3%) | 0.96<br>(-0.3%) | 1.11<br>(-0.3%) | 1.18<br>(+0.2%) | 1.05<br>(-0.3%) |
| Metadata-Only (RF) | 1.07<br>(-52.9%) | 1.05<br>(-22.1%) | 0.89<br>(+7.3%) | 1.08<br>(+2.7%) | 1.17<br>(+0.8%) | 1.39<br>(-32.4%) |
| Metadata-Only (GBM) | 0.61<br>(+12.9%) | 0.78<br>(+9.3%) | 0.82<br>(+14.6%) | 0.99<br>(+10.8%) | 1.08<br>(+8.5%) | 1.06<br>(-1.0%) |
| Full Model (RF) | 0.74<br>(-5.7%) | 0.77<br>(+10.5%) | 0.69<br>(+28.1%) | 0.83<br>(+25.2%) | 0.96<br>(+18.6%) | 0.87<br>(+17.1%) |
| Full Model (GBM) | 0.63<br>(+10.0%) | 0.73<br>(+15.1%) | 0.70<br>(+27.1%) | 0.86<br>(+22.5%) | 0.94<br>(+20.3%) | 0.88<br>(+16.2%) |

*Note.* TF-IDF: Term Frequency-Inverse Document Frequency; RF: Random Forest; GBM: Gradient Boosting Machine. Percentages in parentheses represent change compared to the dummy regressor RMSE. "Full Model" refers to models where both the item meta-data and LLM-extracted features are included as predictors. The Direct LLM Estimation row is included as a cross-approach reference. MAE results (reported in the Appendix Table A1) show a similar pattern.

**Figure 4.**

*Association between Feature-Based Estimates and True Difficulty Estimates by Grade for Math*

Model: Random Forest

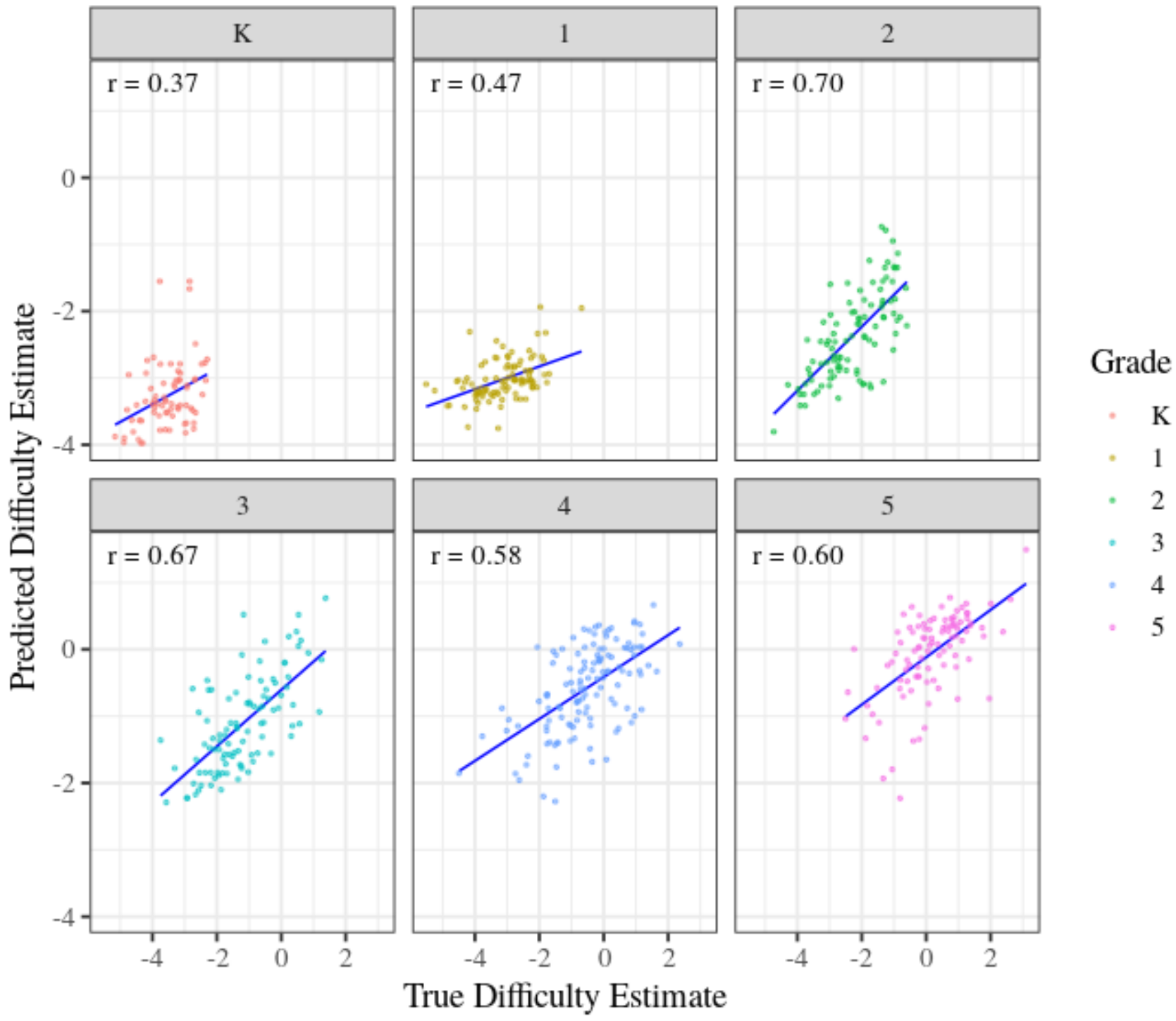

Model: Gradient Boosting Machine (GBM)

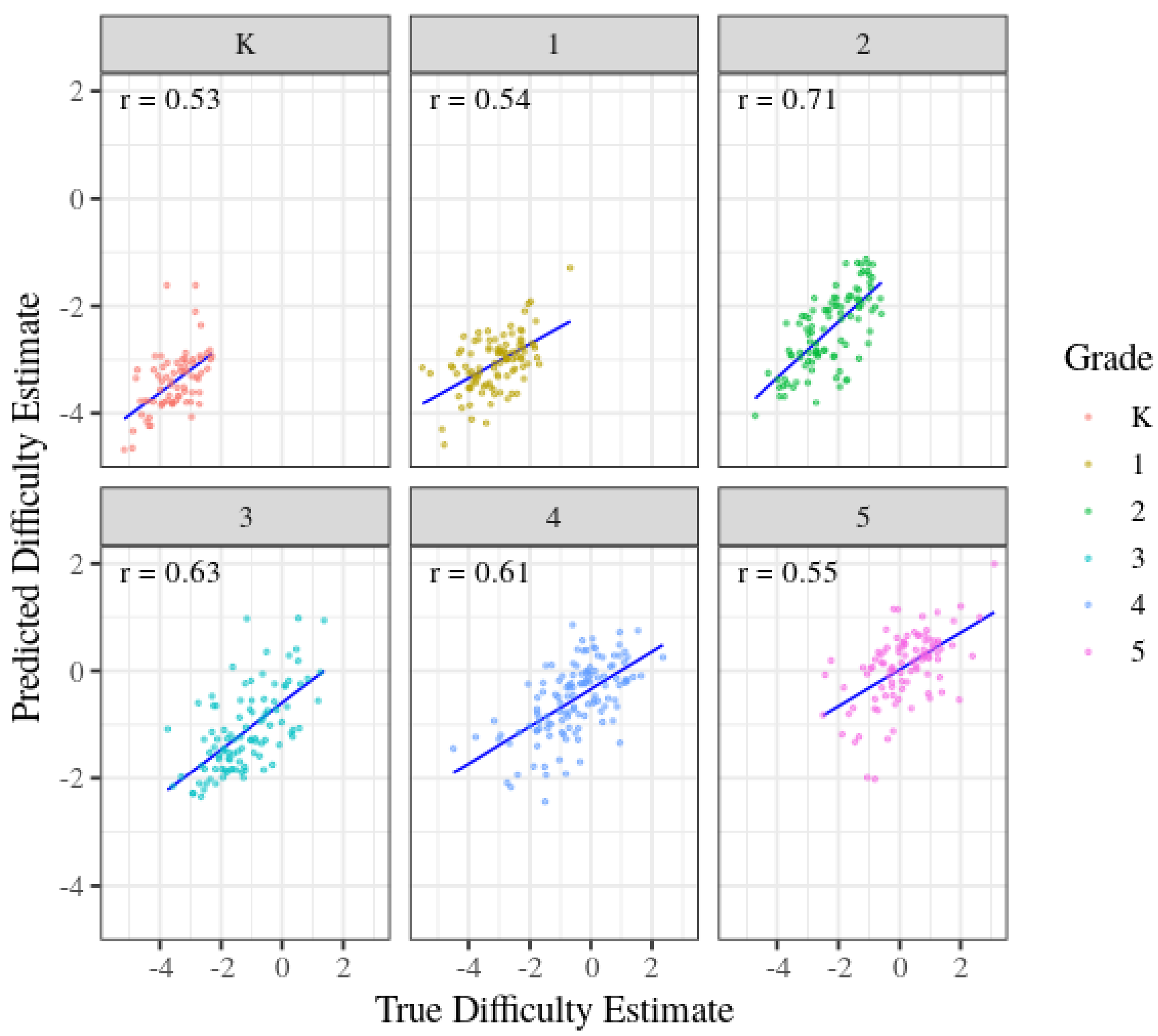

***Reading Predictions Using Random Forest.*** The overall comparison of true difficulty estimates and those predicted by the random forest across all grades produced an RMSE of 0.72 and MAE of 0.58, showing that the model outperforms the direct LLM estimation approach (RMSE = 0.86, MAE = 0.69), the dummy regressor benchmarks (RMSE = 1.04, MAE = 0.84), the TF-IDF + Random Forest benchmark (RMSE = 1.09, MAE = 0.86), and the metadata-only random forest model (RMSE = 0.87, MAE = 0.67). Furthermore, the predicted estimates were highly correlated with the true difficulty estimates ($r$ = .87; Figure 5). By-grade analyses indicate that the model's prediction accuracy is consistently better than the benchmarks and the direct LLM estimation approach for all grades (0.53 ≤ RMSE ≤ 0.89; 0.42 ≤ MAE ≤ 0.72; Table 7). Further, the predicted and true difficulty estimates are highly correlated for all grades (.62 ≤ $r$ ≤ .82; Figure 6).

***Reading Predictions Using GBM.*** The overall model performance for GBM was similar to random forest (RMSE = 0.73, MAE = 0.59, $r$ = .87; Figure 5). Further, by-grade analyses results were consistent with those from the random forest, both in terms of model errors (Table 7) and in terms of correlations between predicted and true item difficulties (Figure 6).

Overall, both the random forest and gradient boosting models demonstrated strong predictive performance across grades. Compared to the dummy regressor, the random forest reduced RMSE by 31.6% on average, and improved upon the metadata-only random forest by 17.6% on average. The gradient boosting model showed comparable gains, reducing RMSE by 31.0% relative to the dummy regressor and by 10.9% relative to the metadata-only gradient boosting baseline.

**Figure 5.**

*Association between Feature-Based Estimates and True Difficulty Estimates for Reading*

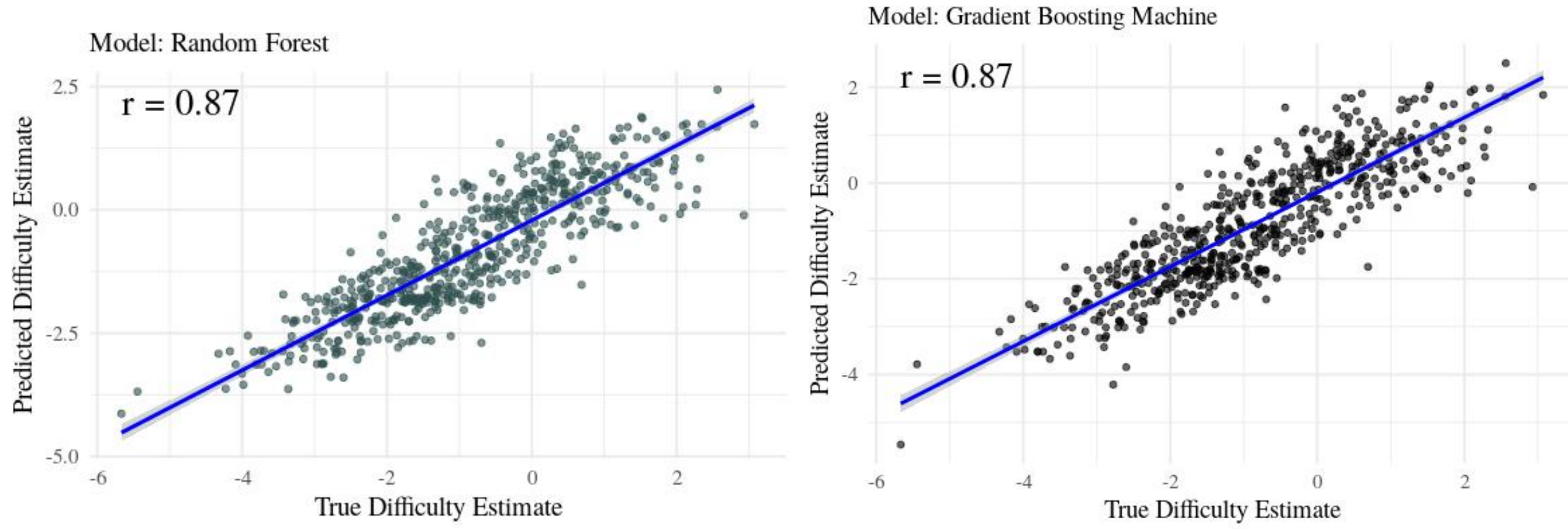

**Table 7.**

*RMSE by Grade Based on the Feature-Based Estimation Approach for Reading*

| | Grade | | | | | |
|---|---|---|---|---|---|---|
| Approach | K | 1 | 2 | 3 | 4 | 5 |
| Dummy Regressor | 0.90 | 0.90 | 1.10 | 1.14 | 1.06 | 1.13 |
| Direct LLM Estimation | 0.82 (+8.9%) | 0.89 (+1.1%) | 0.96 (+12.7%) | 0.78 (+31.6%) | 0.82 (+22.6%) | 0.89 (+21.2%) |
| TF-IDF + RF | 1.17 (-30.3%) | 0.92 (-2.2%) | 1.09 (+1.0%) | 1.14 (-0.4%) | 1.03 (+2.9%) | 1.14 (-1.0%) |
| Metadata-Only (RF) | 0.89 (+1.1%) | 0.75 (+16.7%) | 0.92 (+16.4%) | 0.80 (+29.8%) | 0.86 (+18.9%) | 0.96 (+15.0%) |
| Metadata-Only (GBM) | 0.65 (+27.8%) | 0.71 (+21.1%) | 0.85 (+22.7%) | 0.81 (+28.9%) | 0.84 (+20.8%) | 0.96 (+15.0%) |
| Full Model (RF) | 0.53 (+41.1%) | 0.64 (+28.9%) | 0.78 (+29.1%) | 0.66 (+42.1%) | 0.77 (+27.4%) | 0.89 (+21.2%) |
| Full Model (GBM) | 0.53 (+41.1%) | 0.61 (+32.2%) | 0.79 (+28.2%) | 0.67 (+41.2%) | 0.81 (+23.6%) | 0.91 (+19.5%) |

*Note.* TF-IDF: Term Frequency-Inverse Document Frequency; RF: Random Forest; GBM: Gradient Boosting Machine. Percentages in parentheses represent change compared to the dummy regressor RMSE. "Full Model" refers to models where both the item meta-data and LLM-extracted features are included as predictors. The Direct LLM Estimation row is included as a cross-approach reference. MAE results (reported in the Appendix Table A2) show a similar pattern.

**Figure 6.**

*Association between Feature-Based Estimates and True Difficulty Estimates by Grade for Reading*

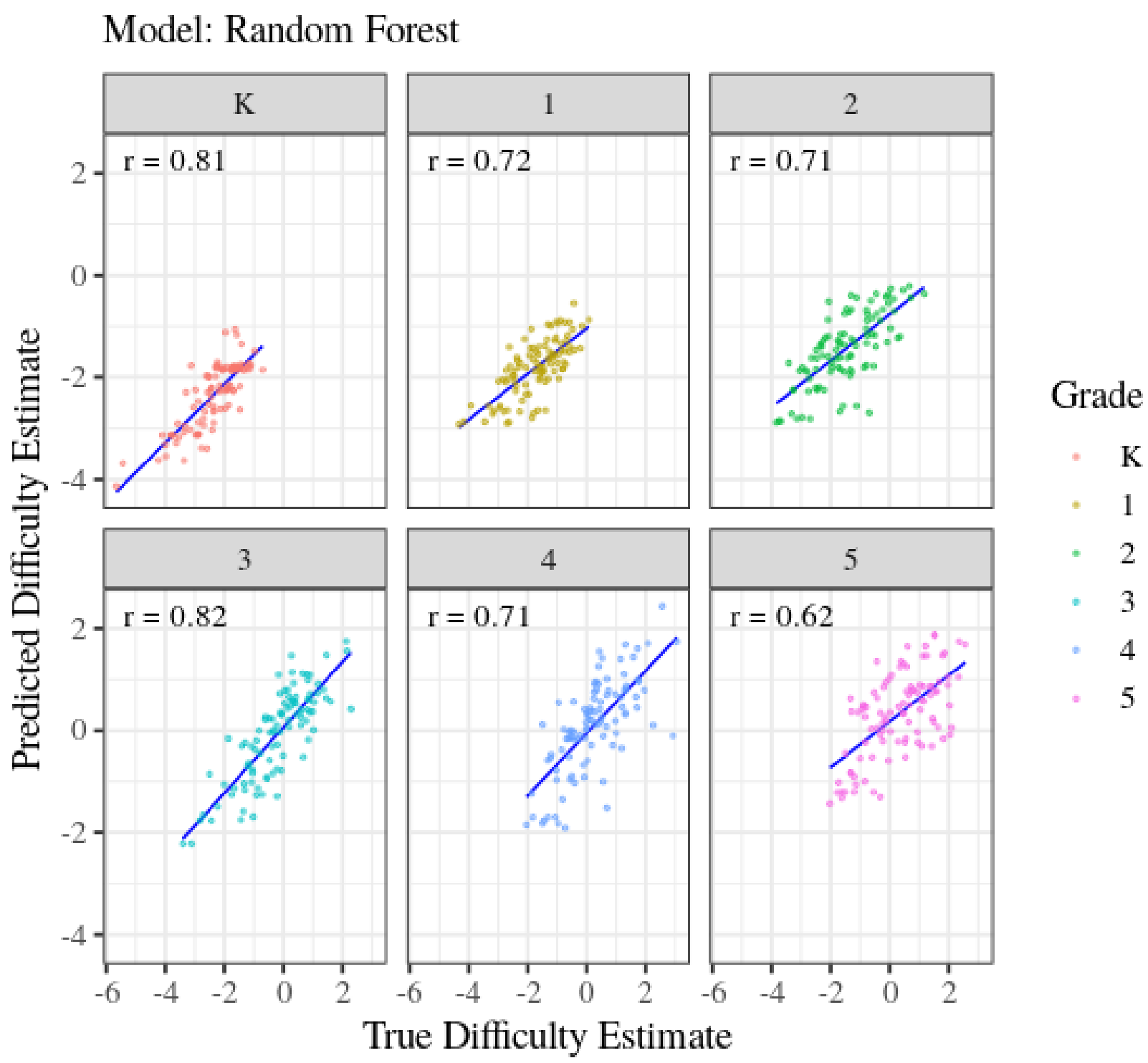

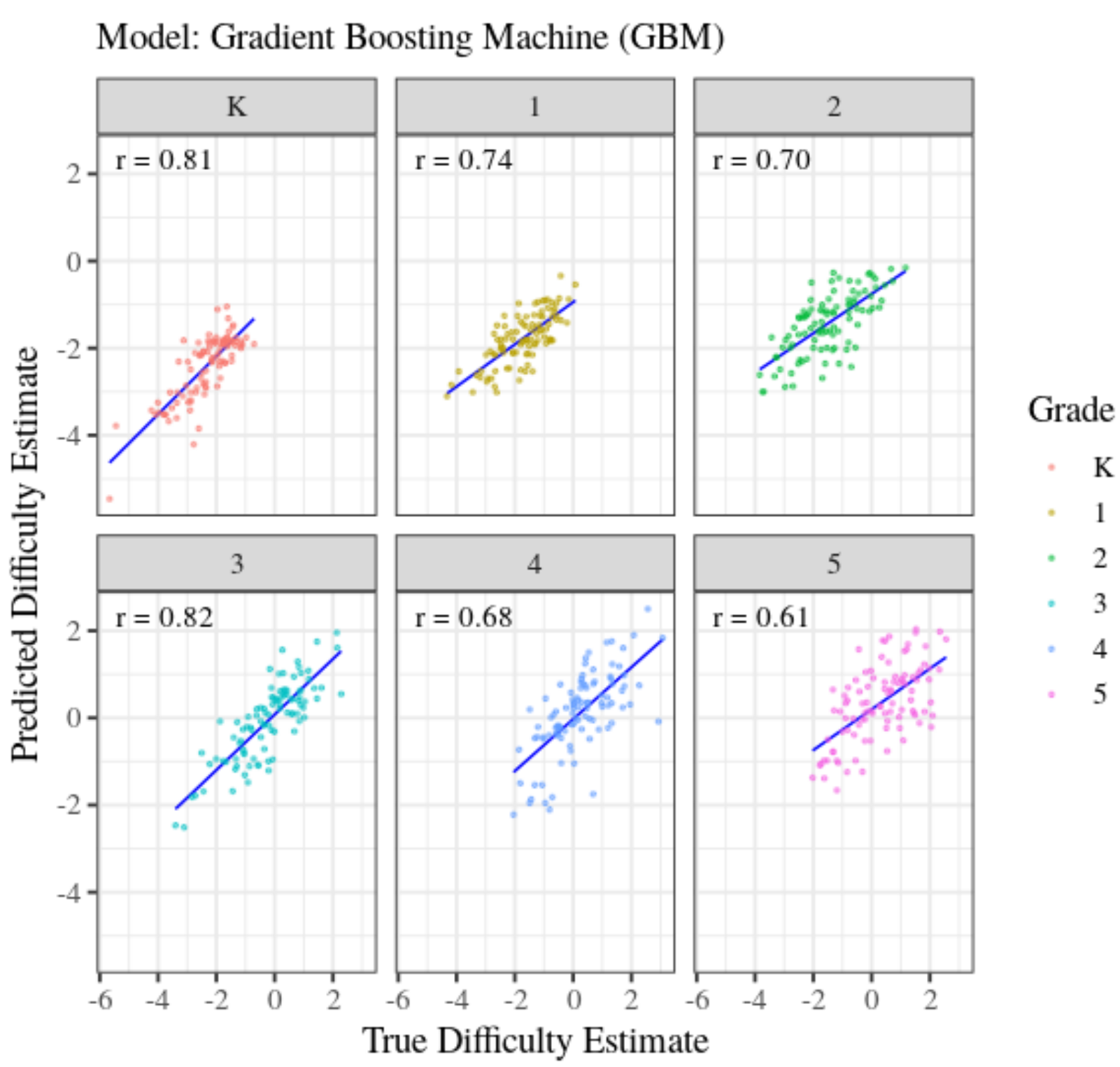

### *Feature Importance*

While the primary objective of this study was predictive accuracy, we conducted post-hoc analyses to interpret which features were most influential in the models' predictions. It is crucial to note that feature importance is model-specific; different algorithms may attribute importance differently based on their internal mechanics. Therefore, the findings should be interpreted as insights into how each specific model leveraged the provided features to estimate difficulty. Accordingly, we explored feature importance using two distinct, model-appropriate techniques: permutation-based importance for the random forest models (Breiman, 2001) and SHAP (SHapley Additive exPlanations; Lundberg & Lee, 2017) for the gradient boosting machines.

***Variable Importance for Random Forests.*** For the random forest models, we assessed feature importance using permutation-based importance, a technique that evaluates the contribution of each feature by measuring how much the model's performance degrades when that feature's information is disrupted. Specifically, the importance score for a given feature is calculated as the percentage increase in the model's Mean Squared Error (MSE) after the values of that feature are randomly shuffled. A larger increase in MSE signifies a greater reliance on that feature for accurate predictions. Table 8 presents the top 20 most important features for the math and reading models. For both subjects, the models' predictive power is not derived from a few dominant variables, but rather from a large and varied set of features that includes both objective item metadata (e.g., Grade Level, Word Count) and the more granular LLM-rated features (e.g., Use of Visuals, Vocabulary Complexity). This indicates that the models are leveraging a rich combination of information to make their predictions.

**Table 8.**

*Feature Importance Rankings for the Top 20 Features in the Random Forest Models*

| Math | | | Reading | | |
|---|---|---|---|---|---|
| Rank | Feature[1] | Importance[2] | Rank | Feature[1] | Importance[2] |
| 1 | M: Grade Level | 33.71 | 1 | M: Word Count | 33.00 |
| 2 | M: Word Count | 29.77 | 2 | M: Grade Level | 28.63 |
| 3 | M: Domain: Numbers & Operations | 26.48 | 3 | M: Domain: Reading Foundations | 27.84 |
| 4 | F: Use of Visuals | 24.70 | 4 | F: Passage-Based Item | 24.68 |
| 5 | F: Student Engagement | 23.10 | 5 | M: Domain: Reading Informational Text | 24.32 |
| 6 | M: Type: Fill-in-the-blank | 21.64 | 6 | F: Student Engagement | 22.44 |
| 7 | F: Distractor Trickiness | 21.48 | 7 | F: Inference Required | 21.37 |
| 8 | F: Text Translation | 20.94 | 8 | F: Vocabulary Complexity | 21.21 |
| 9 | F: Focus: Conceptual Understanding | 20.77 | 9 | M: Domain: Reading Literature | 19.86 |
| 10 | M: Domain: Measurement, Data, & Statistics | 20.31 | 10 | F: Syntax Complexity | 19.72 |
| 11 | M: Type: Multiple Choice | 20.06 | 11 | F: Overall Difficulty (LLM Estimate) | 19.54 |
| 12 | F: Use of Math-Specific Terms | 19.98 | 12 | F: Skill Challenge | 16.98 |
| 13 | F: Skill Challenge | 19.60 | 13 | F: Distractor Trickiness | 16.98 |
| 14 | M: Domain: Geometry | 19.50 | 14 | F: Depth of Knowledge | 16.84 |
| 15 | F: Linguistic Cues | 19.24 | 15 | F: Cognitive Load | 16.83 |
| 16 | F: Cognitive Load | 18.91 | 16 | F: Figurative/Abstract Language | 16.10 |
| 17 | F: Overall Difficulty (LLM Estimate) | 18.31 | 17 | F: Multi-Step Reasoning | 11.11 |
| 18 | F: Real-World Application | 17.24 | 18 | M: Type: Multiple Choice | 9.48 |
| 19 | F: Symbolic Reasoning | 15.76 | 19 | M: Type: Hot Text | 7.22 |
| 20 | F: Depth of Knowledge (DOK) | 14.42 | 20 | M: Type: Graphic Gap Match | 3.44 |

[1] Features starting with “M” (e.g., “M: Grade Level”) are item meta-data. Features starting with "F” (e.g., “F: Use of Visuals”) represent LLM-rated features.
[2] The Importance score represents the percentage increase in the model's Mean Squared Error (MSE) when a single feature's values are randomly permuted. A higher score indicates that the model relies more heavily on that feature to make accurate predictions.

A consistent finding across both subjects is the importance of two metadata features: Grade Level and Word Count. For both math and reading, these simple, objective measures were among the top predictors in the random forest models. Beyond these, the models leveraged a mix of other metadata and LLM-rated features. For math, specific Subject Domains (e.g., Numbers & Operations, Measurement, Data, & Statistics) and Item Types (e.g., Fill-in-the-blank) were highly influential, alongside LLM-rated features like Use of Visuals, Student Engagement, and Distractor Trickiness. For reading, specific domains like Reading Foundations were key predictors, followed by LLM-rated linguistic features such as Presence of a Passage, Inference Required, and Vocabulary Complexity.

***Variable Importance for Gradient Boosting Machines.*** To interpret the gradient boosting models, we used SHapley Additive exPlanations (SHAP), a game-theory-based approach that explains how each feature contributes to an individual prediction (Lundberg et al., 2018; Lundberg & Lee, 2017). The results are visualized in SHAP beeswarm plots for the top 15 features for each subject (Figure 7); we chose to display 15 features to maintain the legibility of the visualization. In these plots, each point represents a single item. The vertical axis lists the features, ordered by their overall importance (mean absolute SHAP value). The horizontal axis shows the SHAP value, indicating how much a feature's value for a specific item shifted the difficulty prediction away from the average. Positive SHAP values pushed the prediction towards higher difficulty, while negative values pushed it towards lower difficulty. The color of each point represents the feature's original value for that item, from low (gold) to high (purple).

Similar to the findings from Random Forests, across both math and reading, we see that Grade Level and Word Count are among the more important features. The plots show a clear and intuitive relationship for these variables: low values (gold dots) are concentrated on the left with negative SHAP values, while high values (purple dots) are on the right with positive SHAP values. This indicates the model learned a strong positive correlation where higher grades and greater word counts reliably push the prediction toward higher difficulty.

For the math model, the analysis highlights the predictive role of the LLM-rated Skill Challenge and the LLM-Estimated Overall Difficulty. As shown in Figure 7, higher ratings on these two features are strongly associated with increased predicted item difficulty. Beyond these top predictors, the model leverages a balanced combination of other metadata and LLM-rated features. For instance, the Numbers & Operations subject domain and the LLM's rating of the possibility of Linguistic Misunderstanding both demonstrate a moderate influence on the model's predictions, illustrating a reliance on both objective item categorization and nuanced cognitive attributes.

The interpretation for the reading model is particularly noteworthy. The plot reveals that the LLM-extracted Syntax Complexity emerged as the single most important feature, outweighing even the metadata predictors like Grade Level and Word Count. Other LLM-rated features, such as Skill Challenge and Tricky Distractors, also rank among the top predictors. Similar to the math model, the remaining feature list confirms that the model's accuracy is derived from a healthy mixture of metadata (such as the Reading Foundations domain) and various LLM-rated features, each contributing in non-trivial ways to the final prediction.

**Figure 7.**

*SHAP Plots for the Gradient Boosting Machines*

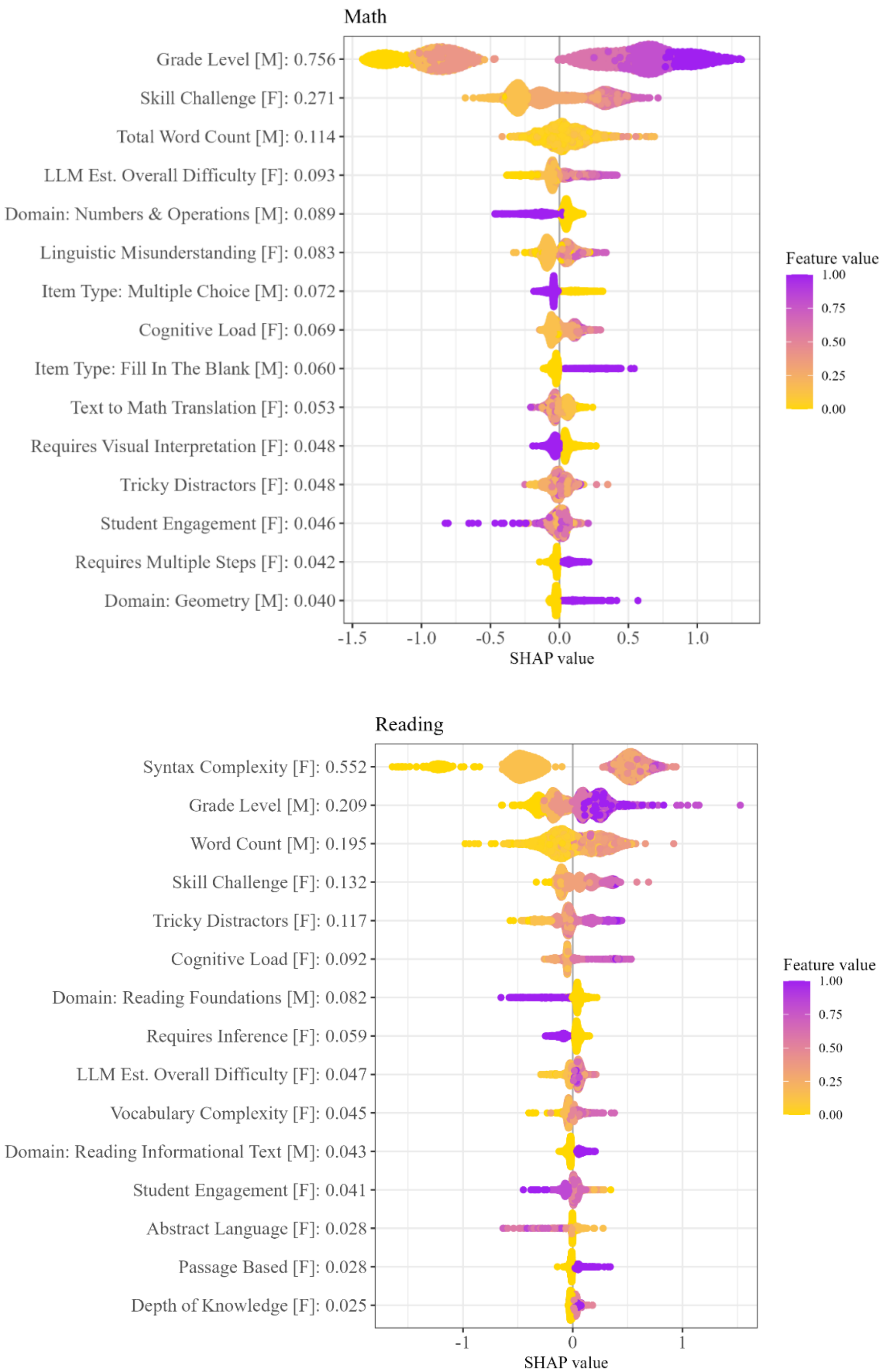

## Discussion

The primary aim of this study was to investigate the capabilities of large language models in estimating K-5 math and reading item difficulties both directly (via zero-shot, text-based prompting) and indirectly (through feature extraction and tree-based modeling). Results overall showed promise for LLM-generated difficulty predictions, while also highlighting certain limitations, particularly with younger grade items, when compared to dummy-regressor benchmarks and prior literature.

A central finding was that GPT-4o's direct difficulty estimates (i.e., a zero-shot numerical rating of item difficulty from 1 to 100) exhibited a moderate to strong correlation with empirically derived Rasch item difficulty parameters when collapsing across grades; however, accuracy varied substantially by grade level. More specifically, the overall correlation between LLM-predicted and actual difficulty for the holdout dataset for math and reading were .83 and .81, respectively. These coefficients suggest that, when considering the entire K-5 set, GPT-4o's direct ratings tracked well with "true" item difficulties. However, when examining each grade individually, the results were more uneven. For math at grades K and 1, the LLM's direct estimates were often no better (and in some cases worse) than a dummy regressor model that simply predicted the average item difficulty. This contrasted with higher correlations and significantly lower error for grades 3, 4, and 5. A similar trend emerged in reading: prediction accuracy was weakest for grades K and 1 but improved sharply from grades 2 through 5. One possible explanation for these disparities is the range restriction of item difficulties in lower grades. The distribution and range of the lower performing grades for math (i.e., grades K and 1) are noticeably lower (average $SD$ = 0.80, average range = 4.62) than the higher grades (average $SD$ = 1.08, average range = 6.56). Similarly, the lower performing grades for reading were more restricted in their distribution and range of item difficulties (average *SD* = 0.89, average range = 5.39) compared to the higher performing grades (average *SD* = 1.13, average range = 6.33). The narrower spread of true item difficulties in lower grades may make it more challenging for the LLM to differentiate among items at the lower end of the difficulty continuum, whereas the broader range and greater variance in higher-grade items provide clearer differentiation and thus bolster predictive accuracy.

By comparison, the feature-based estimation approach where GPT-4o was used to extract specific item features for input into random forest or gradient boosting models yielded stronger overall performance. For math, random forests and gradient boosting both achieved correlations of .87 with the true item difficulties, with average errors (RMSE and MAE) notably below both the direct LLM estimates and the dummy model benchmarks. In reading, ensemble tree-based regressors also outperformed direct LLM ratings (e.g., overall RMSE improved from 0.86 to approximately 0.72–0.73). These differences were especially pronounced for the early-grade items. Notably, both the TF-IDF + Random Forest baseline and the metadata-only models demonstrated that simple text features or metadata alone were insufficient to reach the predictive accuracy of the full LLM-based feature extraction approach. For the present item pool, the TF-IDF baseline performed no better than the dummy regressor, highlighting that traditional keyword-based NLP methods cannot capture the deeper semantic and cognitive features that drive item difficulty. Similarly, while metadata-only models showed some predictive utility (particularly the GBM variant), they underperformed the full models that incorporated LLM-extracted cognitive and linguistic features. This pattern confirms that the LLM's contribution extends well beyond what can be captured through simple metadata or surface-level text analysis.

Taken together, these results suggest that more structured, feature-based methods provide superior predictive accuracy across the full elementary range. The superior performance of the feature-based approach can be attributed to its method of decomposition. The direct estimation method is analogous to asking an expert for a single, holistic difficulty score. This forces the LLM to internally and simultaneously weigh dozens of complex factors and then collapse that intricate reasoning into a single number—a process that can be inconsistent, especially for nuanced items at the lower grade levels. In contrast, the feature-based approach is like providing the LLM with a detailed rubric. By prompting the model to rate a series of discrete, well-defined characteristics (e.g., *Syntax Complexity*, *Cognitive Load*), we break down the complex task of judging "difficulty" into smaller, more manageable, and likely more reliable evaluations.

This two-step process then leverages a second specialist: the ensemble tree-based algorithm. This algorithm's sole task is to learn the optimal weighting of each feature "rubric score" to best predict the true difficulty. It can empirically determine from the data, for example, that Syntax Complexity is highly predictive for reading items while the presence of Visuals is more important for math—a complex weighting that is difficult to achieve in a single, zero-shot judgment. This hybrid approach, therefore, combines the LLM's sophisticated linguistic and cognitive analysis with the statistical power of a machine learning model, resulting in a more robust, transparent, and accurate system.

It is important to note that the two approaches we examined operate at different levels of supervision and therefore provide different types of information. The zero-shot method relies primarily on the language model's internalized linguistic and conceptual knowledge. In contrast, the feature-based approach combines LLM-extracted features with ensemble tree-based algorithms trained on a relatively large set of empirically calibrated items. Thus, while the feature-based models predictably outperform the zero-shot method, the two are not strictly parallel in terms of data and analytic requirements. We view the zero-shot estimates as a relatively lower-effort, minimal-information baseline that demonstrates what can be achieved by leaning on the language model's internal knowledge and reasoning, and the feature-based/tree-based estimates as a semi-supervised alternative that illustrates the accuracy gains possible when LLM insights are integrated with psychometric modeling.

Direct comparison of our findings with those from other studies is challenging. Models are often trained on datasets that differ substantially in terms of subject area, content variability, and the distribution and range of item difficulty (for an overview, see Štěpánek et al., 2023). With these caveats in mind, we believe several comparisons are informative for situating the present results within the broader literature of item difficulty estimation.

First, our feature-based approach can be seen as a conceptual continuation of pre-LLM, feature-based methods. In much of that earlier research, such as the influential work by Ha et al. (2019), a wide range of surface-level and linguistic features—including lexical counts, syntactic structures, and semantic ambiguity indicators—were extracted using traditional NLP pipelines. While important and informative, this work yielded modest predictive performance; their best model, combining all feature types, reported correlations up to $r = .32$ with percent-correct values computed using human-generated item responses. Our work represents both continuity and a significant departure. We build on the tradition of text-based feature extraction, but by leveraging an LLM, we can operationalize constructs at a much deeper level. Instead of relying on indirect proxies, the LLM can directly evaluate higher-order cognitive and semantic aspects like the need for multi-step reasoning, the plausibility of distractors, and implicit inference demands. The substantially stronger correlations achieved in our feature-based models (*r* values frequently exceeding .70) underscore the advantage of this approach to feature engineering.

Second, when compared to other recent studies that specifically employ language models, our results are highly encouraging. Many "AI as examinee" approaches, where LLMs simulate student responses, have reported modest correlations. For example, the zero-shot LLaSA framework achieved correlations up to $r = .35$ (Park et al., 2024), and fine-tuned "AI examinees" yielded correlations between $r \approx .08$ and .22 with IRT difficulty parameters (Maeda, 2025). Other approaches using transformer-based embeddings have shown slightly better performance, with correlations in the range of $r \approx .52$–.59 (McCarthy et al., 2021). Our *direct LLM estimation* approach produced correlations that are generally competitive with or superior to these prior efforts, particularly for reading items. However, our *feature-based* approach, which combines LLM-generated features with ensemble tree models, demonstrates a clear step forward, often yielding correlations that are higher than those previously reported in the LLM-based literature.

Finally, a comparison against benchmark performance further highlights the strength of our models. In a recent, comprehensive effort to estimate the difficulty of clinical multiple-choice questions involving 12 different teams, Yaneva et al. (2024) report the same dummy-regressor benchmark as ours. They note that "even the best solution out-performed the baseline by only a small margin" (p. 473), with an RMSE of 0.299 compared to the dummy-regressor baseline's RMSE of 0.311 (a difference of just 0.012). In contrast, the performance gains in the present study were far more substantial. The difference between the predicted and dummy-regressor RMSEs for the gradient boosting models ranged from 0.06 to 0.26 for math (Table 6) and 0.27 to 0.47 for reading (Table 7). These much larger margins suggest that the models achieved a comparatively stronger and more meaningful performance gain over the baseline, reinforcing the promise of a feature-based difficulty estimation approach that combines the analytical power of LLMs with machine learning algorithms.

Practically, these findings demonstrate a potential path toward faster, more cost-effective item difficulty estimation. For example, providing item developers with a difficulty estimation tool during the item generation process can assist them with targeting general difficulty ranges. Furthermore, traditional item calibration requires administering items to large student samples, leading to delays, concerns about item overexposure, and logistical expense. If LLM-based methods can predict item difficulty at scale with relatively low error rates and high, but imperfect, correlations, these estimates could reduce the need for large sample sizes, as researchers can use a Bayesian approach to incorporate LLM-based difficulty estimates as informative priors.

**Recommended Workflow for Testing Professionals**

Based on the lessons learned throughout this program of research, we provide the following workflow for researchers and practitioners who would want to implement a similar item difficulty estimation approach with a different item pool (see Figure 8):

1. Selecting the items: A sufficiently large sample is needed to support both training and testing of the model. While there are no strict guidelines for determining sample size in tree-based models, larger samples generally yield more stable estimates, whereas smaller samples increase the risk of overfitting. When dividing data into training and holdout sets, ensure that both subsets have similar distributions of key item characteristics relevant to your goal. Among these, item difficulty is the most critical; ideally, the distribution of item difficulty should be comparable across the training and testing datasets.

2. Identifying relevant features: This crucial step is best undertaken in collaboration with SMEs in the target domain. We recommend conducting in-depth discussions (in the form of interviews or focus groups), during which SMEs are asked to (a) describe the factors they consider when designing items to ensure they align with a target difficulty range, and (b) identify item characteristics they would examine when tasked with estimating item difficulty. To be thorough, we recommend reviewing the literature to ensure potentially relevant features are included. Once you identify these features, you can generate prompts that instruct the LLM to evaluate each item based on those features.
3. Selecting the language model: There are different factors that could influence this decision, including model cost (better performing models tend to be more expensive, but the correlation is not perfect and might not apply to every task) and data security (i.e., some model providers save your data for future model training). To inform model selection based on performance at the preliminary stage, we recommend testing multiple models on a subset of your data using a zero-shot difficulty estimation prompt to make a preliminary assessment of their performance.
4. Generating the prompts: Using the features identified in step 2, you can generate your first prompt which instructs the model to evaluate different features based on the item content. For each feature, provide (a) a concise but descriptive statement about what you expect (e.g., "What is the Depth of Knowledge (DOK) level required to answer this item correctly?"), and (b) an exact description of what the output should look like (e.g., "Respond with 1, 2, 3, or 4."). Similar to step 3, we recommend testing the initial prompt with a subset of the training data and evaluating the results to determine whether the model is responding to the prompts as expected. Iteratively refine the prompt prior to running feature extraction for the full training data.

5. Evaluating the LLM responses: Before model training, it is crucial to pre-process the features rated by the LLM. First, each feature should be examined for sufficient variance. Features with little to no variance—for instance, where the LLM assigns the same rating to nearly all items—lack predictive power and should be considered for removal. Furthermore, if, unlike the present research, model explainability is a high priority, it is beneficial to perform a dimensionality reduction analysis, such as a Principal Component Analysis (PCA). This optional step can consolidate redundant features into fewer, more interpretable components, leading to a more parsimonious final model.
6. Training the model(s): There is a wide range of models, including linear regression, random forest, and gradient boosting machines, that can be used to predict item difficulty based on the LLM-extracted features. Many of these algorithms (e.g., tree-based ensemble models) can be optimized using hyperparameter tuning. It is best practice to complete the model comparison and optimization on the training dataset, and select the better performing model(s) at this stage. Note that you can use different indices, such as average error estimates (e.g., RMSE) and correlations between true and predicted difficulty estimates to evaluate each model. For error estimates, calculating benchmarks based on a dummy-regressor model can be highly informative.
7. Validating the model(s): Once you choose the better-performing model(s) based on the training data, apply them to the holdout sample to see how well the model(s) can estimate item difficulty with "unseen" data. Note that a large difference in model performance between training and testing data may indicate overfitting and might be cause for reevaluating the initial model(s).

**Figure 8.**

*A Seven-Step Workflow for Creating a Feature-Based Item Difficulty Estimation Model*

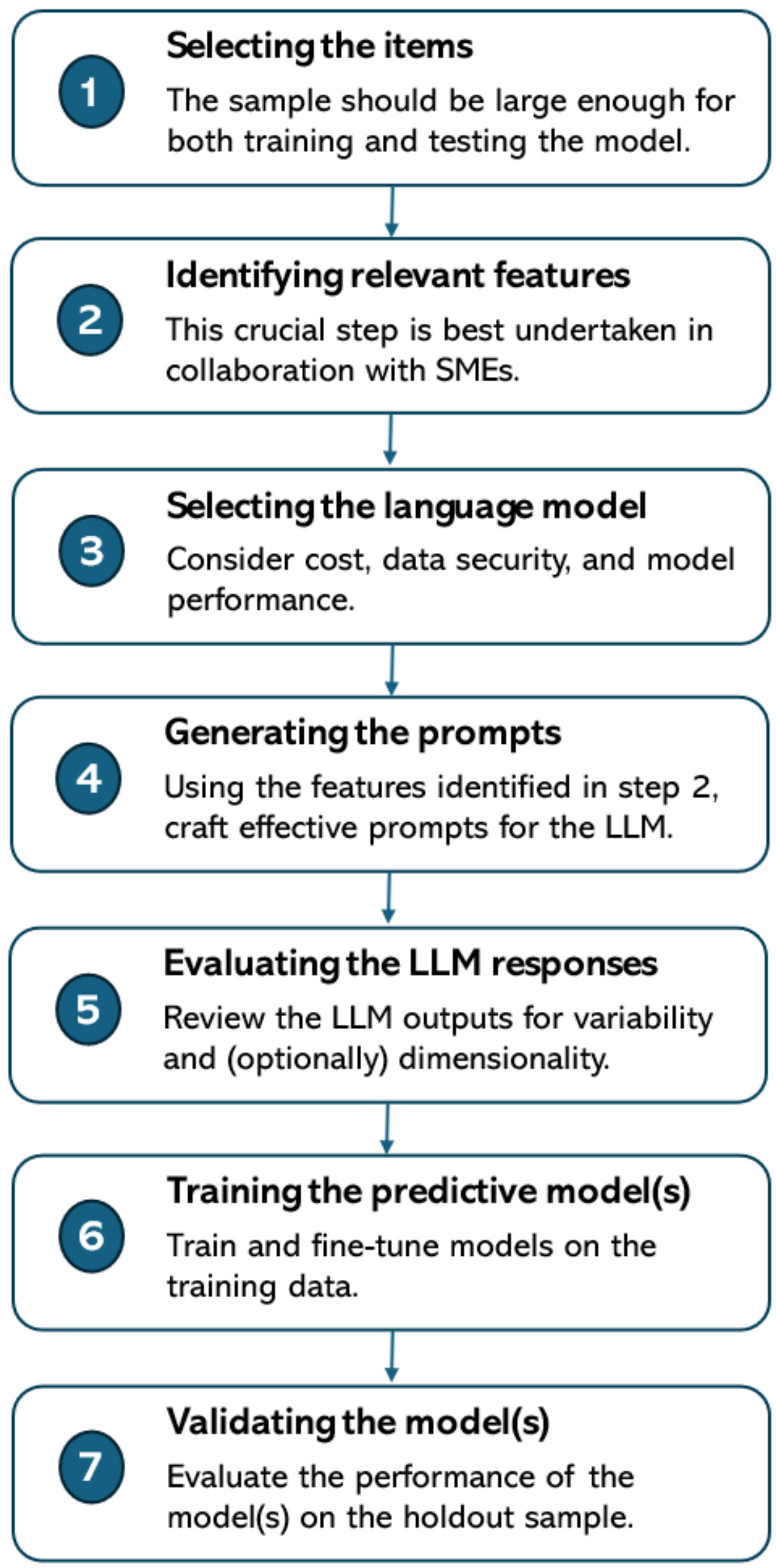

**Limitations and Future Research**

This study targeted K-5 math and reading. While that focus allowed for insights into early-grade item challenges, it remains unclear how well these findings might generalize to higher grade levels or other domains (e.g., science or social studies). Item development traditions vary widely by subject area, and it is possible that LLM-based difficulty estimations would perform differently in disciplines with unique linguistic or conceptual demands.

Another avenue worth considering is the fine-tuning of large language models for item difficulty prediction. While promising in principle, fine-tuning was not feasible in the present study. First, robust fine-tuning requires very large, labeled datasets with calibrated item parameters to avoid overfitting and ensure generalizability. Our dataset of approximately 5,000 items, while large by K–12 assessment standards, is insufficient relative to the scale typically needed for LLM fine-tuning. Second, fine-tuning could raise data security and intellectual property concerns, as assessment items are often proprietary and cannot readily be shared with external model providers. For these reasons, we prioritized prompting- and feature-extraction–based approaches, which are immediately practical for operational use. Future work may revisit fine-tuning as larger item banks become available and as more secure fine-tuning infrastructures for educational data are developed. Furthermore, hybrid approaches that combine feature extraction with structured fine-tuning have shown improved performance in difficulty prediction (e.g., Feng et al., 2025), suggesting a productive avenue for future research.

Our direct LLM estimation approach relied on prompt engineering and refinement. While this design was intended to reflect real-world constraints (i.e., not having access to large, labeled datasets for fine-tuning) future research could investigate whether fine-tuning improves performance. Relatedly, although the sample of more than 5,000 items is relatively large, coverage per grade varies, and certain item types or content areas within each subject may be underrepresented. Future research should evaluate whether the grade-level predictions improve if the machine learning models are trained on much larger samples of items per grade.

Finally, some degree of output variability is inherent when working with large language models. To minimize variability, we used structured prompts with constrained response formats, provided explicit descriptions for each feature, and set the temperature parameter to 0 for all model calls, a configuration that reduces inconsistency in LLM outputs. In addition, the use of large item samples for model training helps mitigate the impact of any residual variability by emphasizing stable patterns across many items rather than single estimates. Nevertheless, future research could more directly quantify LLM response stability by re-prompting items multiple times, particularly in settings with smaller item pools or where fine-grained item-level precision is critical.

**Conclusion**

In this study, we evaluated an LLM's capabilities as a tool for estimating the difficulty of elementary-level math and reading items. Although direct LLM estimates alone were often moderately predictive, a feature-based approach that harnesses an LLM's ability to extract detailed cognitive and linguistic attributes consistently outperformed simpler methods. These findings align with emerging research illustrating the promise of LLMs in test development, yet also highlight that success depends on thoughtful prompt design, specialized feature extraction, and careful modeling. Going forward, LLM-based difficulty estimation has the potential to reduce reliance on expensive field testing and to expedite item development cycles. However, practitioners must remain mindful of limitations around very early grade items and the need for strong psychometric oversight. By refining and expanding on these approaches, researchers and testing professionals can move closer to a robust, scalable framework for automatically predicting item difficulties and supporting more efficient assessment design.

## Statements and Declarations

### *Funding*

The authors did not receive a specific grant for this research from any funding agency in the public, commercial, or not-for-profit sectors. The study was conceived, executed, and prepared as part of the authors' employment at Edmentum.

### *Competing Interests*

Both authors were employees of Edmentum at the time this research was conducted. The study utilizes proprietary data from Edmentum's item pool. The authors declare that this relationship did not influence the study design, analyses, interpretation, or decision to publish.

### *Author Contributions*

All authors contributed to the study conception and design. Material and data preparation and analysis were performed by Pooya Razavi. The first draft of the manuscript was written by Pooya Razavi and reviewed and revised by Sonya Powers. Both authors contributed to the revisions, and read and approved the final manuscript.

### *Data Availability Statement*

The analyses in this study were conducted using a subset of Edmentum’s proprietary item pool. As these assessment items are the intellectual property of Edmentum, they cannot be shared publicly. However, the dataset generated by the large language models (LLMs) based on these items—along with the R code used to implement the regression and machine learning models—are publicly accessible at: https://bit.ly/4jl9b9U.

# Appendix A

## LLM Prompts for the Direct Difficulty Estimation Approach

### The Math Prompt

```
You are an expert in K to 12 Math assessment. Below, you are provided
with item content (i.e., prompt, response options) and meta data
(i.e., grade level, item type) for a specific item.

Think step-by-step using all this information to determine how
difficult or easy this item is. When evaluating this item’s difficulty
level, consider how challenging or advanced the skill assessed by the
item is, how tricky the distractors are, and the cognitive load of
answering the item.

Additionally, when assessing the item difficulty, consider the depth
of knowledge (DOK) required to answer the item correctly. DOK level1
items often involve basic calculations, recall of facts, and use of
formulas or procedures. DOK level 1 tasks often require little to no
cognitive effort beyond remembering or applying a well-known process.
DOK Level 2 often involves applying skills or concepts to solve
problems that require more than one step. Tasks in this level may
require organizing data, making decisions on how to approach the
problem, or interpreting a graph or chart. DOK Level 3 often involves
reasoning, planning, and using evidence to explain or justify
mathematical conclusions. Tasks at this level demand higher-order
thinking, such as analyzing data, constructing models, or solving non-
routine problems. DOK Level 4 often involves extended thinking,
complex reasoning and investigation over time, often involving
multiple concepts or real-world applications. This is the most complex
level, where the solutions require connecting and synthesizing ideas,
making decisions, and justifying solutions in an extended format. Note
that the higher the DOK level, the more complex and challenging the
items usually are.

In addition to DOK, consider other factors that can impact difficulty
or complexity. For example, here are some factors that can make math
items more challenging or complex, and therefore increase item
difficulty: items requiring translation from text to a mathematical
process add complexity, misunderstanding linguistic cues in the
problem stem (e.g., “this number is subtracted from that”) can lead to
errors (which increases item difficulty), situations requiring
students to interpret and apply mathematical concepts to realistic
scenarios increase difficulty, improper fractions and mixed numbers
are more challenging due to additional steps, such as converting mixed
```

numbers to improper fractions, representing or interpreting information using visuals like tables, graphs, or charts can add cognitive load, items requiring multi-step reasoning in stems or choices increase complexity, tasks requiring integration of multiple concepts (e.g., patterns leading to functions) can elevate difficulty, asking students to evaluate others' calculations or reasoning tends to be challenging. Items requiring multiple calculations or intermediate steps (e.g., converting units before solving) increase difficulty.

Use all this information and provide a numerical estimate of item difficulty between 1 (a very easy item, assessing a lower grade skill, involving lower DOK, with lower complexity and less challenging distractors) and 100 (a very difficult item, requiring knowledge of higher grade level skills, involving higher DOK, with high cognitive demand, or tricky distractors). Use the full 1 to 100 range, with the following suggested breakdown: Very easy: 1-19, Easy: 20-39, Moderate: 40-59, Challenging: 60-79, and Very challenging: 80-100.

The Item Contents are provided below. Use all available item information to inform your estimate. Only provide a difficulty estimate (as a 2-decimal number between 1 to 100) and nothing else.

{The full item content is provided here.}

**The Reading Prompt**

You are an expert in K to 12 Reading assessment. Below, you are provided with item content (i.e., prompt, response options) and meta data (i.e., grade level, item type) for a specific item.

Think step-by-step using all this information to determine how difficult or easy this item is. When evaluating this item's difficulty level, consider how challenging or advanced the skill assessed by the item is, how complex the vocabulary is, how tricky the distractors are, and the cognitive load of answering the item.

Additionally, when assessing the item difficulty, consider the depth of knowledge (DOK) required to answer the item correctly. DOK level1 items often involve recall and reproduction. Items at this level often involve basic recall of facts, definitions, or procedures. DOK Level 2 often involves skills and concepts. This level requires applying learned skills and concepts to complete tasks or solve problems. DOK Level 3 often involves strategic thinking. Tasks at this level demand reasoning, planning, and evidence-based decision-making. DOK Level 4 often involves extended thinking. This is the most complex level, and involves extended and integrated reasoning over time. Note that the higher the DOK level, the more complex and challenging the items usually are.

In addition to DOK, also consider other factors that can impact difficulty or complexity. For example, here are some factors that can make reading items more challenging or complex, and therefore increase item difficulty: Items requiring students to infer themes or concepts not explicitly stated in the text are more challenging. Items that assess more than one construct (e.g., identifying a theme and explaining how it's developed) increase complexity. Lack of explicit information in the stem (e.g., not providing the theme) adds cognitive steps and difficulty. Straightforward selection (e.g., identifying the theme) tends to be easier. Tasks requiring multiple responses (e.g., dropdown menus or selecting both a theme and supporting detail) add complexity due to increased error potential. The difficulty of the passage itself (e.g., vocabulary, sentence structure, content) heavily influences item difficulty.

Use all this information and provide a numerical estimate of item difficulty between 1 (a very easy item, assessing a lower grade skill, involving lower DOK, with simple vocabulary and few distractors) and 100 (a very difficult item, requiring knowledge of higher grade level skills, involving higher DOK, with complex vocabulary, high cognitive demand, or tricky distractors). Use the full 1 to 100 range, with the following suggested breakdown: Very easy: 1-19, Easy: 20-39, Moderate: 40-59, Challenging: 60-79, and Very challenging: 80-100.

The Item Contents are provided below. Use all available item information to inform your estimate. Only provide a difficulty estimate (as a 2-decimal number between 1 to 100) and nothing else.

{The full item content is provided here.}

# Appendix B

## LLM Prompts for the Feature Based Difficulty Estimation Approach

### The Math Prompt

```
You are an expert in K-12 Math assessment design and evaluation.
Below, you will find item content (e.g., prompt, stem, response
options) and item metadata (e.g., grade level, item type) about a
specific item.


Please carefully review all the information provided about the item.
Then, respond to the following 20 questions about the item. For
questions requiring a numerical response, provide only the number (1
to 10 or 1 to 4, etc.). For questions requiring a 'Yes' or 'No'
response, respond with 'Y' or 'N'. Do not include any explanations or
additional information beyond the specific response requested.


(1) Rate how challenging or advanced the skill assessed by the item is
on a scale from 1 (very easy) to 10 (very challenging).

(2) On a scale from 1 (not at all) to 10 (a lot), rate the extent to
which correctly answering the item requires students to translate text
to a mathematical process.

(3) How tricky are the distractors (i.e., incorrect response options)?
Rate on a scale from 1 to 10.

(4) How demanding is the item in terms of cognitive load? In your
response consider working memory and executive function load (e.g.,
How many pieces of information, like partial sums, sub-results, must
be held in mind or manipulated?). Rate on a scale from 1 to 10.

(5) What is the Depth of Knowledge (DOK) level required to answer this
item correctly? Respond with 1, 2, 3, or 4.

(6) Misunderstanding linguistic cues in the problem stem (e.g., “this
number is subtracted from that”) can lead to errors. How likely are
the students in this grade to misunderstand the linguistic cues in the
problem (1: not at all to 10: a lot)?

(7) Does the item involve situations requiring students to interpret
and apply mathematical concepts to realistic or real-world scenarios?
'Y' or 'N'.
```

(8) Does the item include improper fractions and mixed numbers, or require the students to convert mixed numbers to improper fractions? 'Y' or 'N'.

(9) Does answering the question involve representing or interpreting information using visuals (e.g., tables, graphs, or charts)? 'Y' or 'N'.

(10) Does the item require multi-step reasoning? 'Y' or 'N'.

(11) Does the item involve integrating multiple concepts (e.g., patterns leading to functions)? 'Y' or 'N'.

(12) Does the item ask students to evaluate others' calculations or reasoning? 'Y' or 'N'.

(13) Does solving the problem require multiple calculations or intermediate steps (e.g., converting units before solving)? 'Y' or 'N'.

(14) How engaging is the item for a typical student in the target grade? Rate from 1 (not engaging) to 10 (extremely engaging).

(15) Does the item require symbolic reasoning (e.g., basic algebraic expressions or variable manipulation)? 'Y' or 'N'.

(16) Does the item use math-specific terms (e.g., "product," "quotient," "divisor," "factor," "variable")? 'Y' or 'N'.

(17) Does the item have multi-part instructions (e.g., "First do X, then do Y, then fill in Z")? 'Y' or 'N'.

(18) Does solving the item correctly require repeated calculations and multiple sub-conclusions? 'Y' or 'N'.

(19) Some items hinge on conceptual understanding (e.g., place value or fraction sense), others on arithmetic accuracy. Which best describes this item: 1 = Conceptual understanding, 2 = Arithmetic accuracy, 3 = Both, 4 = Neither.

(20) Considering all relevant factors and information, provide a numerical estimate of item difficulty between 1 (a very easy item, assessing a lower grade skill, involving lower DOK, with lower complexity and less challenging distractors) and 100 (a very difficult item, requiring knowledge of higher grade level skills, involving higher DOK, with high cognitive demand, or tricky distractors). Use the full 1 to 100 range, with the following suggested breakdown: Very easy: 1-19, Easy: 20-39, Moderate: 40-59, Challenging: 60-79, and Very challenging: 80-100.

```
Use all the item information below to inform your responses:


{The full item content is provided here.}


Return your final answers in the following JSON format only (and no
extra text--do not wrap your JSON with triple backticks or add
additional words like 'json' in your response):

{
  "F1_SkillChallenge": "X",
  "F2_TextToMath": "X",
  "F3_DistractorTricky": "X",
  "F4_CognitiveLoad": "X",
  "F5_DOK": "X",
  "F6_LinguisticMisunderstanding": "X",
  "F7_RealWorld": "X",
  "F8_ImproperFractions": "X",
  "F9_VisualInterpretation": "X",
  "F10_MultiStep": "X",
  "F11_MultipleConcepts": "X",
  "F12_EvaluateOthers": "X",
  "F13_MultiCalculations": "X",
  "F14_Engagement": "X",
  "F15_SymbolicReasoning": "X",
  "F16_MathTerms": "X",
  "F17_MultiPartInstructions": "X",
  "F18_RepeatedCalcs": "X",
  "F19_ItemType": "X",
  "F20_OverallDifficulty": "X"
}
```

**The Reading Prompt**

```
You are an expert in K-12 Reading assessment design and evaluation.
Below, you will find item content (e.g., prompt, stem, response
options) and item metadata (e.g., grade level, item type) about a
specific item.


Please carefully review all the information provided about the item.
Then, respond to the following 13 questions about the item. For
questions requiring a numerical response, provide only the number (1
to 10 or 1 to 4, etc.). For questions requiring a 'Yes' or 'No'
response, respond with 'Y' or 'N'. Do not include any explanations or
additional information beyond the specific response requested.


(1) Rate how challenging or advanced the skill assessed by the item is
on a scale from 1 (very easy) to 10 (very challenging).

(2) Rate the complexity or advancement of the vocabulary used in the
item on a scale from 1 to 10.

(3) How tricky are the distractors (i.e., incorrect response options)?
Rate on a scale from 1 to 10.

(4) How demanding is the item in terms of cognitive load? In your
response consider working memory and executive function load (e.g.,
How many pieces of information must be held in mind or manipulated?)
as well as metacognitive requirements (e.g., Does the item require
students to self-monitor, reflect on reading strategies, or evaluate
multiple claims?). Rate on a scale from 1 to 10.

(5) What is the Depth of Knowledge (DOK) level required to answer this
item correctly? Respond with 1, 2, 3, or 4.

(6) Does the item require students to infer themes or concepts that
are not explicitly stated in the text? Respond with 'Y' or 'N'.

(7) Does the item assess more than one construct (e.g., identifying a
theme and explaining how it’s developed)? Respond with 'Y' or 'N'.

(8) On a scale from 1 to 10, how would you rate the syntax complexity
of the item (e.g., based on the presence of compound or complex
sentences, embedded clauses, or passive voice)?

(9) How would you rate the use of figurative or abstract language in
this item? Respond on a scale from 1 (none) to 10 (heavy use).
```

(10) How engaging is the item for a typical student in the target grade? Rate on a scale from 1 (not engaging) to 10 (extremely engaging).

(11) If the item has a passage, rate the passage's difficulty on a scale from 1 to 10. If there is no passage, respond with 'NA'.

(12) Does answering the question require multiple steps or multi-part reasoning (e.g., "First identify the main idea, then find evidence to support it, then explain the author's purpose.")? Respond with 'Y' or 'N'.

(13) Considering all relevant factors and information, provide a numerical estimate of item difficulty between 1 (a very easy item, assessing a lower grade skill, involving lower DOK, with lower complexity and less challenging distractors) and 100 (a very difficult item, requiring knowledge of higher grade level skills, involving higher DOK, with high cognitive demand, or tricky distractors). Use the full 1 to 100 range, with the following suggested breakdown: Very easy: 1-19, Easy: 20-39, Moderate: 40-59, Challenging: 60-79, and Very challenging: 80-100.

Use all the item information below to inform your responses:

{The full item content is provided here.}

Return your final answers in the following JSON format only (and no extra text--do not wrap your JSON with triple backticks or add additional words like 'json' in your response):

```
{{
  "F1_SkillChallenge": "X",
  "F2_VocabComplexity": "X",
  "F3_DistractorTricky": "X",
  "F4_CognitiveLoad": "X",
  "F5_DOK": "X",
  "F6_RequiresInference": "X",
  "F7_MultipleConstructs": "X",
  "F8_SyntaxComplexity": "X",
  "F9_AbstractLanguage": "X",
  "F10_Engagement": "X",
  "F11_PassageDifficulty": "X",
  "F12_MultipleSteps": "X",
  "F13_OverallDifficulty": "X"
}}
```

## Appendix C

## Supplementary Tables

**Table A1.**

*MAE by Grade for Math Across Direct and Feature-Based Estimation Approach*

| | Grade | | | | | |
|---|---|---|---|---|---|---|
| Approach | K | 1 | 2 | 3 | 4 | 5 |
| Dummy Regressor | 0.58 | 0.70 | 0.82 | 0.91 | 0.92 | 0.83 |
| Direct LLM Estimation | 0.60 (-3.4%) | 0.69 (+1.4%) | 0.71 (+13.4%) | 0.78 (+14.3%) | 0.76 (+17.4%) | 0.75 (+9.6%) |
| TF-IDF + RF | 0.58 (0.0%) | 0.69 (+1.4%) | 0.82 (0.0%) | 0.87 (+4.4%) | 0.93 (-1.1%) | 0.82 (+1.2%) |
| Metadata-Only (RF) | 0.89 (-53.4%) | 0.82 (-17.1%) | 0.75 (+8.5%) | 0.84 (+7.7%) | 0.94 (-2.2%) | 1.14 (-37.3%) |
| Metadata-Only (GBM) | 0.49 (+15.5%) | 0.62 (+11.4%) | 0.67 (+18.3%) | 0.76 (+16.5%) | 0.84 (+8.7%) | 0.82 (+1.2%) |
| Full Model (RF) | 0.59 (-1.7%) | 0.61 (+12.9%) | 0.55 (+32.9%) | 0.64 (+29.7%) | 0.73 (+20.7%) | 0.69 (+16.9%) |
| Full Model (GBM) | 0.49 (+15.5%) | 0.56 (+20.0%) | 0.54 (+34.1%) | 0.67 (+26.4%) | 0.73 (+20.7%) | 0.70 (+15.7%) |

*Note.* TF-IDF: Term Frequency-Inverse Document Frequency; RF: Random Forest; GBM: Gradient Boosting Machine. Percentages in parentheses represent percent change in MAE relative to the dummy regressor for the same grade. Positive values indicate lower MAE than the dummy regressor, whereas negative values indicate higher MAE. “Full Model” refers to models where both the item meta-data and LLM-extracted features are included as predictors.

**Table A2.**

*MAE by Grade for Reading Across Direct and Feature-Based Estimation Approach*

| | Grade | | | | | |
|---|---|---|---|---|---|---|
| Approach | K | 1 | 2 | 3 | 4 | 5 |
| Dummy Regressor | 0.69 | 0.74 | 0.90 | 0.90 | 0.84 | 0.96 |
| Direct LLM Estimation | 0.63 (+8.7%) | 0.73 (+1.4%) | 0.79 (+12.2%) | 0.59 (+34.4%) | 0.64 (+23.8%) | 0.73 (+24.0%) |
| TF-IDF + RF | 0.84 (-21.7%) | 0.76 (-2.7%) | 0.88 (+2.2%) | 0.91 (-1.1%) | 0.81 (+3.6%) | 0.97 (-1.0%) |
| Metadata-Only (RF) | 0.63 (+8.7%) | 0.59 (+20.3%) | 0.76 (+15.6%) | 0.65 (+27.8%) | 0.63 (+25.0%) | 0.78 (+18.7%) |
| Metadata-Only (GBM) | 0.46 (+33.3%) | 0.57 (+23.0%) | 0.69 (+23.3%) | 0.65 (+27.8%) | 0.64 (+23.8%) | 0.80 (+16.7%) |
| Full Model (RF) | 0.42 (+39.1%) | 0.53 (+28.4%) | 0.65 (+27.8%) | 0.55 (+38.9%) | 0.58 (+31.0%) | 0.72 (+25.0%) |
| Full Model (GBM) | 0.42 (+39.1%) | 0.51 (+31.1%) | 0.66 (+26.7%) | 0.56 (+37.8%) | 0.62 (+26.2%) | 0.74 (+22.9%) |

*Note.* TF-IDF: Term Frequency-Inverse Document Frequency; RF: Random Forest; GBM: Gradient Boosting Machine. Percentages in parentheses represent percent change in MAE relative to the dummy regressor for the same grade. Positive values indicate lower MAE than the dummy regressor, whereas negative values indicate higher MAE. “Full Model” refers to models where both the item meta-data and LLM-extracted features are included as predictors.